\documentclass[twocolumn,showpacs,nofootinbib,superscriptaddress]{revtex4-1}

\usepackage[nointlimits]{amsmath}
\usepackage{hyperref}
\usepackage{amsfonts}
\usepackage{amssymb}
\usepackage{amsmath}
\usepackage{amsthm}
\usepackage{latexsym}
\usepackage{graphicx}
\usepackage{float}
\usepackage{color}
\usepackage{layout}
\usepackage[english]{babel}

\usepackage{wrapfig}

\def\LL{\mathcal{L}}

\def\OO{{\cal O}}

\begin{document}

\title{Cassini and extra force constraints to nonminimally coupled gravity with screening mechanism}

\author{Riccardo March}
\email{r.march@iac.cnr.it}
\affiliation{Istituto per le Applicazioni del Calcolo, CNR, Via dei Taurini 19, 00185 Roma, Italy}
\affiliation{INFN - Laboratori Nazionali di Frascati (LNF), Via E. Fermi 40, Frascati 00044 Roma, Italy}

\author{Orfeu Bertolami}
\email{orfeu.bertolami@fc.up.pt}
\affiliation{Departamento de F\'isica e Astronomia,\\Faculdade de Ci\^encias da Universidade do Porto, Rua do Campo Alegre 687, 4169-007 , Porto, Portugal}
\affiliation{Centro de F\'isica das Universidades do Minho e do Porto, Rua do Campo Alegre 687, 4169-007, Porto, Portugal}

\author{Marco Muccino}
\email{Marco.Muccino@lnf.infn.it}
\affiliation{INFN - Laboratori Nazionali di Frascati (LNF), Via E. Fermi 40, Frascati 00044 Roma, Italy}

\author{Cl\'audio Gomes}
\email{claudio.gomes@fc.up.pt}
\affiliation{Departamento das Ci\^encias da F\'isica, Qu\'imica e Engenharia, Faculdade de Ci\^encias e Tecnologia da Universidade dos A\c{c}ores,\\ Campus de Ponta Delgada, Rua da M\~ae de Deus 9500-321 Ponta Delgada, Portugal}
\affiliation{Centro de F\'isica das Universidades do Minho e do Porto, Rua do Campo Alegre 687, 4169-007, Porto, Portugal}

\author{Simone Dell'Agnello}
\email{simone.dellagnello@lnf.infn.it}
\affiliation{INFN - Laboratori Nazionali di Frascati (LNF), Via E. Fermi 40, Frascati 00044 Roma, Italy}

\date{today}

\begin{abstract}
We consider a nonminimally coupled curvature-matter gravity theory at the Solar System scale. 
Both a fifth force of Yukawa type and a further non-Newtonian extra force that arises from the nonminimal coupling are present in the solar
interior and in the solar atmosphere up to interplanetary space. The extra force depends on the spatial gradient of space-time curvature $R$.
The conditions under which the effects of such forces can be screened by the chameleon mechanism and be made consistent with Cassini measurement 
of PPN parameter $\gamma$ are examined. 
This consistency analysis requires a specific study of Sun's dynamical contribution to the arising forces at all its layers.
\end{abstract}

\maketitle

\section{Introduction}

Cosmological and astrophysical phenomena, such as the accelerated expansion of the Universe and the flattening of the galaxy rotation curves,
can be explained by resorting to dark energy and dark matter, respectively. Nevertheless, such phenomena could arise, in principle, from a
modification of General Relativity (GR) at astrophysical (galactic and extragalactic) and cosmological scales. Among such modifications,
$f(R)$ gravity \cite{Capoz-1,Carroll,Capoz-2,DeFTs} involves the replacement of the Ricci curvature scalar $R$, in the Einstein-Hilbert action, 
by a nonlinear function $f(R)$.

A further modification of GR is nonminimally (NMC) coupled gravity, where the Einstein-Hilbert action is replaced with a more general form
involving two functions of curvature $f^1(R)$ and $f^2(R)$ \cite{BBHL}. The function $f^1(R)$ has a role analogous to $f(R)$ gravity theory, and the function
$f^2(R)$ multiplies the matter Lagrangian density giving rise to a nonminimal coupling between geometry and matter. This possibility has been extensively studied in the context of dark matter \cite{drkmattgal}, dark energy \cite{curraccel}, inflation \cite{c}, energy density fluctuations \cite{d}, gravitational waves \cite{e}, cosmic virial theorem \cite{f}, Jeans' instability and star formation in Bok globules \cite{g}. This model has also been examined with the Newton-Schrodinger approach \cite{h,i}.

In a previous paper \cite{MPBD} the case of functions $f^1(R),f^2(R)$ analytic at $R=0$ was considered, and
constraints to the resulting NMC gravity model have been computed through perturbations to perihelion precession
by using data from observations of Mercury's orbit. 

It turns out that NMC gravity modifies the gravitational attraction
by introducing both a fifth force of the Yukawa type and an extra force which depends on the spatial gradient of the Ricci scalar $R$.
While the Yukawa force is typical also of $f(R)$ gravity, the existence of the extra force is specific of NMC gravity \cite{BBHL,BLP},
and it is an effect of the nonminimal coupling that induces a non-vanishing covariant derivative of the energy-momentum tensor. The arising Yukawa contribution can give origin to static solutions even though in the absence of pressure \cite{i}.

In Ref. \cite{MBMBD} constraints to the NMC gravity model with analytic $f^1,f^2$ functions have been computed by using the results
of a geophysical experiment which looked for deviations from Newton's inverse square law in the ocean \cite{Zum}.
It turns out that the presence of the extra force in a fluid such as seawater imposes more stringent constraints on the NMC gravity model
than the observation of both Mercury's perihelion precession and lunar geodetic precession. 
Hence for the NMC gravity model with analytic functions, because of the extra force, 
Solar System constraints are weaker than geophysical constraints.

In order to look for meaningful Solar System constraints to NMC gravity, in the present paper we consider the case of a function $f^2(R)$
which contains a term proportional to $R^\alpha$, with $\alpha<0$, so that $f^2(R)$ is not analytic at $R=0$.
The resulting model has been used in Ref. \cite{drkmattgal} to predict the flattening of the galaxy rotation curves,
and to predict the current accelerated expansion of the Universe \cite{curraccel}.

Since $f^2(R)$ is not analytic, the method based on the $1/c$ expansion used in Ref. \cite{MPBD} does not work for the present model.
A completely different nonlinear method has to be used.
It turns out that in Solar System the above NMC model exhibits a screening mechanism, which is the NMC version of the so called {\it chameleon mechanism}
\cite{KW}, which makes it possible to compute Solar System constraints.

In the present paper we adapt to NMC gravity the nonlinear computations made in Ref. \cite{HS} for
the chameleon mechanism in the gravity field of the Sun, assuming spherical symmetry.
The nonminimal coupling complicates the computation with respect to the case of $f(R)$ gravity considered in Ref. \cite{HS},
though the solution turns out to have essentially the same general qualitative properties.
We compute an analytic approximation of the chameleon solution of the field equations and we find
constraints on the parameters of the NMC gravity model from the Cassini measurement of PPN parameter $\gamma$ \cite{Cassini}.
In order to satisfy the Cassini constraint the chameleon solution turns out to be close to GR inside a \emph{screening radius} $r_s$ that has to be large enough,
particularly, $r_s$ either lies inside the solar convection zone, close to the top of the zone, or it is larger. Deviations from GR are sourced by the fraction of solar
mass, including solar atmosphere, contained in the region with radii $r>r_s$, so that if $r_s$ lies in the convection zone then such deviations are essentially sourced
by a \emph{thin shell} of mass in the convection zone, which is a typical property of the chameleon mechanism \cite{KW}.

Moreover, by computing the equations of hydrodynamics in the solar interior and atmosphere, both modeled as a perfect fluid, we find
the expressions of the fifth force of Yukawa type and of the non-Newtonian extra force which is a consequence of the nonminimal coupling.
The shape of both forces is affected by the screening mechanism and depends on the dynamical contribution of the various Sun's layers.
Then we compute the effect of the NMC extra force that is expected to be relevant where the mass density gradient is large: for instance, 
close to the Sun's edge in the solar atmosphere, particularly in the chromosphere and
in the transition region between the chromosphere and the inner solar corona. Further constraints on the gravity model are then computed
by resorting to spectroscopic observations of the solar atmosphere. 

Before performing our computation, let us point out that in an interesting recent work \cite{Fisher-Carlson} it has been argued that for NMC models where the coupling function, $f^2(R)$, is dominated by negative powers of the scalar curvature, one should expect sizeable effects of small curvatures at very dense nuclear physics situations, which demands for a considerable suppression. This is clearly a quite special situation, as negative scalar curvature effects were designed to have implications in astrophysical and cosmological situations. Negative powers models also lead to pathological situations for black holes for which $R=0$. In fact, the $f^2(R)$ function of the NMC curvature-matter models were not devised to have a single universal term, but to be a sum of terms that would be relevant at specific scales. Thus, the point made by of Ref. \cite{Fisher-Carlson} must be regarded as the statement that at nuclear physics scales the NMC effects are suppressed. In what concerns the calculation that will be carried out in the present work, this issue can be addressed through the proposed screening mechanism which would yield a plausible scenario for the desired suppression of the extra force inside a nucleus. Indeed, the nucleus has to be unscreened, hence the screening radius is zero, due to its small mass. Thus, with the expression of the extra force outside the screening radius computed in this paper, the expression for the density gradient is multiplied by a negative power of density. In other words, the screening mechanism would suppress the extra force inside the nucleus where the density is large. 

The paper is organized as follows. In Section II the NMC gravity model is specified. In Section III we compute an approximate solution of the field equations inside and around the Sun,
namely the chameleon solution. In Section IV we compute the equations of hydrodynamics of a perfect fluid (under the assumption of spherical symmetry)
and we find the expressions of the fifth force and extra force in the fluid; particularly, we find an approximation of the extra force in regions of the solar atmosphere
where the density gradient is large. In Section V we compute the constraints from Cassini measurement of $\gamma$. In Section VI we compute the effect
of the extra force in the chromosphere-corona transition region an we look for constraints from spectroscopic measurements.
Conclusions are drawn in Section 7. A model of mass density profile, both for the solar interior and the atmosphere, is reported in an appendix and used
to find analytical order of magnitude estimates of the constraints on the parameters of the gravity model.

\section{Nonminimally coupled gravity}

In the present work we consider gravitational theories with an action functional of the form \cite{BBHL},
\begin{equation}\label{action-funct}
S = \int \left[\frac{1}{2}f^1(R) + [1 + f^2(R)] \LL_m \right]\sqrt{-g} \, d^4x,
\end{equation}
where $f^i(R)$ (with $i=1,2$) are functions of the Ricci scalar curvature $R$, $\LL_m$ is the Lagrangian
density of matter, and $g$ is the metric determinant.
The standard Einstein-Hilbert action of GR is recovered by taking
\begin{equation}
f^1(R) = \frac{c^4}{8\pi G}R, \qquad f^2(R) = 0,
\end{equation}
where $G$ is Newton's gravitational constant.

The variation of the action functional with respect to the metric $g_{\mu\nu}$ yields the field equations:
\begin{eqnarray}\label{field-eqs}
& &\left(f^1_R + 2f^2_R \LL_m \right) R_{\mu\nu} - \frac{1}{2} f^1 g_{\mu\nu}  \\
& &= \left(\nabla_\mu \nabla_\nu -g_{\mu\nu} \square \right) \left(f^1_R + 2f^2_R \LL_m \right)
+ \left(1 + f^2 \right) T_{\mu\nu}, \nonumber
\end{eqnarray}
where $f^i_R \equiv df^i\slash dR$. The trace of the field equations is given by
\begin{eqnarray}\label{trace}
& &\left( f^1_R + 2f^2_R \LL_m \right) R - 2f^1 + 3\square f^1_R +
6\square\left( f^2_R \LL_m \right)  \nonumber\\
& &= \left( 1 + f^2 \right) T,
\end{eqnarray}
where $T$ is the trace of the energy-momentum tensor $T_{\mu\nu}$.

A relevant feature of NMC gravity is that the energy-momentum tensor of matter is not covariantly conserved, indeed, applying the Bianchi identities to Eq. (\ref{field-eqs}), one finds that
\begin{equation}\label{covar-div-1}
\nabla_\mu T^{\mu\nu} = \frac{f^2_R }{ 1 + f^2} ( g^{\mu\nu} \LL_m - T^{\mu\nu} ) \nabla_\mu R,
\end{equation}
a result that, as discussed thoroughly in Refs. \cite{multiscalar,Sotiriou1}, cannot be ``gauged away'' by a convenient conformal transformation, but is instead a distinctive feature of the model under scrutiny. 

This property, when applied to the hydrostatic equilibrium of the Sun, will play an important role in constraining NMC gravity.

\subsection{Metric and energy-momentum tensors}

We use the following notation for indices of tensors:
Greek letters denote space-time indices ranging from 0 to 3,
whereas Latin letters denote spatial indices ranging from 1 to 3.
The signature of the metric tensor is $(-,+,+,+)$.

The Sun is modelled as a stationary distribution of matter with spherical symmetry.
Then the metric which describes the spacetime in the gravitational field of the Sun, and in the solar neighbourhood of the galaxy,
has the general spherically symmetric isotropic form around a source centred at $r=0$:
\begin{eqnarray} \label{metric}
ds^2= &-& \left[1-2\Phi(r)+2\Psi(r)\right]c^2dt^2 \nonumber\\
&+& \left[1+2\Phi(r)\right] (dr^2 + r^2 d\Omega^2),
\end{eqnarray}
where the potentials $\Phi$ and $\Psi$ are perturbations of the Minkowski metric such that $\left|\Phi(r)\right| \ll 1$ and $\left|\Psi(r)\right| \ll 1$.
For the purpose of the present paper the functions $\Phi$ and $\Psi$ will be computed at order $\OO(1/c^2)$. These considerations are consistent with previous assumptions to tackle the hydrostatic equilibrium in the relativistic limit \cite{stellequil}.

The components of the energy-momentum tensor in spherical coordinates, to the relevant order for our computations and
in the case of spherical symmetry and radial motion, are given by (Ref. \cite{Wi}, Chapter 4.1):
\begin{eqnarray}
T^{tt} &=& \rho c^2 + \OO\left(1\right), \qquad T^{tr} = \rho c v + \OO\left(\frac{1}{c}\right), \label{T-tt-tr}\\
T^{rr} &=& \rho v^2 + p+ \OO\left(\frac{1}{c^2}\right), \label{T-rr}\\
T^{\theta\theta} &=& \frac{p}{r^2} + \OO\left(\frac{1}{c^2}\right), \label{T-theta-theta}\\
T^{\varphi\varphi} &=& \frac{p}{r^2\sin^2\theta} + \OO\left(\frac{1}{c^2}\right), \label{T-phi-phi}\\
T^{t\theta }&=& T^{t\varphi} = T^{r\theta} = T^{r\varphi} = T^{\theta\varphi} = 0, \label{T-other-mixed}
\end{eqnarray}
where matter (Sun's interior and solar atmosphere) is considered as a perfect fluid with matter density $\rho$, velocity $v$, and pressure $p$.

The Sun is assumed in hydrostatic equilibrium with the exception of the outer solar corona where the dynamical equilibrium of a steady
atmosphere is considered in order to take into account solar wind \cite{Stix}. Density and pressure are radial functions $\rho=\rho(r),p=p(r)$
and velocity $v=v(r)$ in the outer corona is also radial.

The trace of the energy-momentum tensor is
\begin{equation}
T = -\rho c^2 + \OO\left(1\right).
\end{equation}
In the present paper we use $\LL_m = -\rho c^2+\OO(1)$ for the Lagrangian density of matter \cite{BLP}.

\subsection{Choice of functions $f^1(R)$ and $f^2(R)$}

Part of computations will be made for general functions $f^1(R)$ and $f^2(R)$, while the constraints to NMC gravity will be exploited
for the following specific choice of functions:
\begin{equation}\label{f1-f2-specific}
f^1(R) = \frac{c^4}{8\pi G} R, \qquad f^2(R) = q_1R + q_2R^\alpha,\quad \alpha<1,
\end{equation}
where the function $f^1(R)$ corresponds to GR and $q_1,q_2$ and $\alpha$ are real numbers that have to be considered as parameters of the NMC model of gravity.

The functions (\ref{f1-f2-specific}), with $q_1=0$ and negative values of the exponent $\alpha$, have been used in Ref. \cite{drkmattgal} to model the rotation curves of galaxies,
and in Ref. \cite{curraccel} to model the current accelerated expansion of the Universe. The case $q_2=0$ and $q_1\neq 0$ has been used in Ref. \cite{stellequil}
to model stellar equilibrium with computation of solar observables. If both terms with $q_1$ and $q_2$ are present, then in Refs. \cite{drkmattgal,curraccel}
the authors conjecture, for $\alpha<0$, that the term with coefficient $q_2$ dominates at large distances and low densities, while the term with $q_1$
dominates inside astrophysical objects with high densities such as the Sun.

In the present paper we consider the simultaneous presence of both terms and we find, for negative $\alpha$, that the term in $f^2(R)$ with
coefficient $q_1$ dominates in the Sun's interior, while the parameter $q_2$ becomes important in the constraint
by Cassini measurement. The results are in agreement with the above conjecture.

\section{Approximate solution of the field equations}

We begin by adapting to NMC gravity the method of solution applied in \cite{HS} to $f(R)$ gravity. 
Since  the metric potentials $\Phi$ and $\Psi$ are small perturbations we neglect 
the higher order terms that include products of potentials or their derivatives, and cross-products between their derivatives and the potentials.
By computing the Ricci tensor and curvature it follows that the functions $\Phi$ and $\Psi$ satisfy the following equations \cite{HS}:
\begin{eqnarray}
& &\nabla^2(\Phi+\Psi) = -\frac{R}{2}, \label{Phi-Psi-equation}\\
& &\nabla^2\Psi = -\frac{1}{2}\left( R^0_0 + \frac{R}{2} \right). \label{Psi-equation}
\end{eqnarray}
We introduce the scalar field $\eta$ which is a function of curvature $R$ also explicitly depending on the radial coordinate $r$:
\begin{equation}\label{eta-definition}
\eta = \eta(r,R) = f^1_R -2f^2_R \, \rho(r)c^2.
\end{equation}
Aiming to consider the stationary case we have
\begin{equation}
\square\eta = \left(1-2\Phi\right)\frac{d^2\eta}{dr^2} + \left(\frac{2}{r}-\frac{4}{r}\Phi + \frac{d\Psi}{dr}\right)\frac{d\eta}{dr},
\end{equation}
and neglecting the products of $\Phi$ and $d\Psi\slash dr$ by the derivatives of $\eta$,
which are second order quantities, we get $\square\eta \approx \nabla^2\eta$.

In this approximation the time-time component of the field equations (\ref{field-eqs}) is given by
\begin{equation}
\eta R_{00} + \frac{1}{2}\, f^1 = \nabla^2\eta +(1+f^2)\rho c^2,
\end{equation}
and the trace (\ref{trace}) of the field equations becomes
\begin{equation}\label{eta-trace}
\eta R + 3\nabla^2\eta -2f^1 = -(1+f^2)\rho c^2.
\end{equation}
Following Ref. \cite{HS} and combining the time-time component with the trace of the field equations we obtain
\begin{equation}
R^0_0 = \frac{1}{6\eta} \left[ 2\eta R - f^1 -4(1+f^2)\rho c^2 \right],
\end{equation}
then Eq. (\ref{Psi-equation}) for the potential $\Psi$ becomes
\begin{equation}\label{Psi-equation-eta}
\nabla^2\Psi = -\frac{1}{12\eta}\left[ 5\eta R -f^1 -4(1+f^2)\rho c^2 \right].
\end{equation}
In the sequel the functions $f^1(R)$ and $f^2(R)$ are required to satisfy the following conditions in the gravitational field of the Sun
that will have to be verified a posteriori:
\begin{equation}\label{cond-f1-f2}
\left\vert \frac{8\pi G}{c^4} \, \frac{f^1}{R} - 1 \right\vert \ll 1, \qquad \left\vert f^2 \right\vert \ll 1,
\end{equation}
and the following condition on the derivatives of $f^1$ and $f^2$ with respect to $R$,
\begin{equation}\label{cond-f1R-f2R}
\left\vert \frac{8\pi G}{c^4}\eta -1 \right\vert \ll 1.
\end{equation}
The conditions (\ref{cond-f1-f2}) mean that the Lagrangian density in Eq. (\ref{action-funct}) is
a small perturbation of the Lagrangian of GR, while
the condition (\ref{cond-f1R-f2R}), in the case $f^2(R)=0$, becomes a condition used in Ref. \cite{HS} for $f(R)$ gravity theory.
Using such conditions the equations (\ref{Phi-Psi-equation}) and (\ref{Psi-equation-eta}) for $\Phi$ and $\Psi$ are approximately given by:
\begin{eqnarray}
\nabla^2\Phi &=& -\frac{4\pi G}{c^2}\rho + \frac{1}{6}\left( \frac{8\pi G}{c^2}\rho - R \right), \label{Phi-equation-approx} \\
\nabla^2\Psi &=& \frac{1}{3}\left( \frac{8\pi G}{c^2}\rho - R \right), \label{Psi-equation-approx}
\end{eqnarray}
and the trace (\ref{eta-trace}) of the field equations is approximated by
\begin{equation}\label{trace-approx}
\nabla^2\eta = \frac{c^4}{24\pi G}\, R - \frac{1}{3}\rho c^2.
\end{equation}
Note that the above equations are formally the same as the one found in Ref. \cite{HS} for $f(R)$ gravity, with the difference that
the scalar field $\eta$, defined in (\ref{eta-definition}), depends explicitly on $r$ through the multiplication
by $\rho(r)$ due to the nonminimal coupling. Such dependence on $\rho(r)$ will be exploited in the sequel.

\subsection{High-curvature solution}

Following the approach commonly used for this kind of problem \cite{HS,KW}, we introduce
a potential function $V=V(r,\eta)$ and an effective potential $V_{\rm eff}=V-\rho\eta c^2\slash 3$ such that
\begin{equation}\label{eta-equation-potential}
\nabla^2 \eta=\frac{\partial V_{\rm eff}}{\partial\eta}, \qquad \frac{\partial V}{\partial\eta}= \frac{c^4}{24\pi G}\,\omega(\eta,\rho),
\end{equation}
where the function $\omega(\eta,\rho)$ is obtained by solving the equation (\ref{eta-definition}) with respect to $R$. Here we
assume that such a solution exists and it is unique: particularly, this property is satisfied for the specific choice of functions
$f^1(R),f^2(R)$ defined in Eq. (\ref{f1-f2-specific}). Again, the difference with respect to $f(R)$ gravity consists in the explicit dependence
of $\partial V\slash \partial\eta$ on $\rho(r)$ due to the nonminimal coupling.

The effective potential has an extremum which corresponds to the GR solution
\begin{equation}\label{high-curv-sol}
R=\omega(\eta,\rho) = \frac{8\pi G}{c^2}\rho,
\end{equation}
which will be called the high-curvature solution as in Ref. \cite{HS}. We require that such an extremum is a minimum
(see Refs. \cite{HS,KW} and the discussion in the sequel), which yields the condition
\begin{equation}\label{minimum-cond}
\frac{\partial^2 V_{\rm eff}}{\partial\eta^2} = \frac{c^4}{24\pi G}\,\frac{1}{\eta_R} \geq 0,
\end{equation}
with $\eta_R = f^1_{RR}-2f^2_{RR}\rho c^2$ and $R=\omega(\eta,\rho)$, 
the double subscript in $f^i_{RR}$ denoting second derivative with respect to $R$.

{\it Case of specific choice of $f^1,f^2$}. 
The minimum condition (\ref{minimum-cond}) is equivalent to $\eta_R\geq 0$ and,
for the specific choice (\ref{f1-f2-specific}) of functions of curvature, the condition becomes
\begin{equation}
\alpha(\alpha-1)q_2[\omega(\eta,\rho)]^{\alpha-2} \leq 0.
\end{equation}
In the following we assume $q_2\neq 0$, $\alpha\neq 0$ and $\alpha<1$, and the curvature $R=\omega(\eta,\rho)$
positive in the Solar System, so that the minimum condition requires
\begin{equation}\label{q2-sign}
\begin{cases}
0<\alpha<1 & \implies q_2>0, \\
\alpha <0 & \implies q_2<0.
\end{cases}
\end{equation}

\subsubsection{Consistency condition in the Sun's interior}

We require the solution $\eta$ of the trace equation to be a perturbation of the GR solution
in part of the Sun's interior, where we then look for an approximation of the high-curvature solution.
Such a requirement must be met in order to satisfy the constraint from Cassini measurement of PPN parameter $\gamma$.
Moreover, the finiteness of $\nabla^2\eta$ at the origin (Sun's center) imposes the boundary condition
\begin{equation}\label{bound-cond-0}
\frac{d\eta}{dr} = 0, \quad\mbox{at } r=0.
\end{equation}
The high-curvature solution (\ref{high-curv-sol}) is an exact solution of the equation (\ref{trace-approx}) for $\eta$ only
if $\nabla^2\eta=0$, hence if $\eta$ is a harmonic function. Under spherical symmetry, the only harmonic function which satisfies
the boundary condition (\ref{bound-cond-0}) is a constant and, by definition (\ref{eta-definition}) of the function $\eta$, it follows that
the Sun's density would satisfy the condition
\begin{equation}
f^1_R -2f^2_R \, \rho(r)c^2 = constant.
\end{equation}
For instance, for the choice (\ref{f1-f2-specific}) of functions $f^1(R),f^2(R)$ it follows that the density
$\rho(r)$ must also be constant, which is not the case for the Sun's interior. Hence the high-curvature solution can only be
an approximate solution of Eq. (\ref{trace-approx}), and a {\it consistency} condition for such a solution is then
\begin{equation}\label{necess-cond}
\left\vert \nabla^2\left(\eta(r,R=8\pi G\rho(r)\slash c^2)\right) \right\vert \ll \frac{1}{3}\rho c^2.
\end{equation}
Computing the Laplacian of $\eta$ according to (\ref{eta-definition}), and setting at the minimum of $V_{\rm eff}$,
\begin{equation}\label{lambda-definition}
\frac{\partial^2 V_{\rm eff}}{\partial\eta^2} = \frac{1}{\lambda^2} >0, \quad \mbox{for}
\quad \omega(\eta,\rho) = \frac{8\pi G}{c^2}\rho,
\end{equation}
where $\lambda=\lambda(\rho)>0$ has dimension of length and depends on density,
the consistency condition (\ref{necess-cond}) for the high-curvature solution reads
\begin{equation}\label{necess-cond-expl}
\left\vert (\lambda^2-6 f^2_R) \nabla^2\rho + \frac{8\pi G}{c^2}\left( \frac{d\lambda^2}{dR} - 12 f^2_{RR} \right) \left(\frac{d\rho}{dr}\right)^2 \right\vert \ll \rho,
\end{equation}
with $R=8\pi G\rho\slash c^2$.

{\it Case of specific choice of $f^1,f^2$}.
For the specific choice (\ref{f1-f2-specific}) of functions of curvature we have
\begin{equation}\label{lambda-expression}
\lambda^2 = 6 q_2 \alpha(1-\alpha) \left(\frac{8\pi G}{c^2}\rho\right)^{\alpha-1},
\end{equation}
and the consistency condition Eq. (\ref{necess-cond-expl}) reads
\begin{equation}\label{consist-condition}
\left\vert \lambda^2 \left[ \frac{\alpha}{1-\alpha}\nabla^2\rho - \frac{\alpha+1}{\rho}\left(\frac{d\rho}{dr}\right)^2 \right] +6q_1\nabla^2\rho \right\vert \ll \rho.
\end{equation}
If $\alpha<0$ and the order of magnitude of $|\alpha|$ is unity, and if the following conditions hold separately:
\begin{equation}\label{consist-decomposed}
\begin{cases}
\lambda^2\vert \nabla^2\rho \vert & \ll \rho, \\
\lambda\vert d\rho\slash dr\vert & \ll \rho, \\
\vert q_1\nabla^2\rho\vert & \ll \rho,
\end{cases}
\end{equation}
then the consistency condition is satisfied. If $f^2(R)=0$ NMC gravity reduces to $f(R)$ gravity, and in this case
the first two conditions of (\ref{consist-decomposed}) correspond to the consistency condition found in Ref. \cite{HS}.

The explicit expression of the solution for $\eta$ in the Sun's interior will be found in Section \ref{sec:eta-sol-interior}.

\subsubsection{Solution for $\eta$ in the outskirts of the Solar System}\label{Sec:outskirts}

The outer coronal atmosphere of the Sun escapes supersonically into interstellar space giving rise to the solar wind \cite{Stix}.
We approximate the galactic mass density in the solar neighbourhood of the Milky Way with the constant value $\rho_g \approx 10^{-24} \,{\rm g}/{\rm cm}^3$.
We denote by $r_g$ a distance from the Sun's center such that mass density is dominated by the galactic density component for $r>r_g$.
We choose $r_g$ at the heliopause, the boundary between the solar wind and the interstellar medium,
corresponding to a heliocentric radial distance of about $120\mbox{ AU }= 2.58\times 10^ 4R_\odot$, where $R_\odot$ is the Sun's radius.
By using the density model of the outer solar corona in Section \ref{sec:outer-corona} of Appendix B, 
we find for the electron density of the solar wind, at distance $r_g$, the value $n_e\approx 3.24\times 10 ^{-4}\,{\rm cm}^{-3}$ 
not far from that measured by the Voyager spacecrafts \cite{Voyag}. Across the heliopause a large (factor of 20 to 50) density increase takes place
so that mass density reaches the galactic value \cite{Voyag}.

We will assume that in the solar neighbourhood of the galaxy,
for $r>r_g$, the spacetime curvature $R$ is approximately given by the GR solution, $R_g=8\pi G \rho_g /c^2$. 
Hence, for the gravitational field of the galaxy the high-curvature solution holds in the solar vicinity for $r>r_g$, and the field $\eta$
approximately minimizes the effective potential $V_{\rm eff}$. The meaning of such an assumption will be discussed later.

If the condition $R=R_g$ is exactly satisfied for $r>r_g$, then
Eq. (\ref{trace-approx}) implies that $\eta$ is a harmonic function so that, under spherical symmetry, $\eta(r)=a_1+a_2/r$,
with $a_1,a_2$ suitable constants, and by definition (\ref{eta-definition}) of $\eta$,
\begin{equation}
\eta(r) = f^1_{R_g} -2f^2_{R_g} \, \rho_g c^2 = constant, \quad\mbox{for } r>r_g,
\end{equation}
and $a_2=0$. Then, integrating Eq. (\ref{trace-approx}) over spheres of radii $r\geq r_g$ centred at $r=0$,
and using the divergence theorem, we find
\begin{equation}\label{GR-average}
\int_0^{r_g}\left( R - \frac{8\pi G}{c^2}\,\rho \right)r^2 dr = 0.
\end{equation}
However, we will find at the end of Section \ref{sec:screen-convect}
that the presence of the gravitational field of the Sun makes the integral in (\ref{GR-average}) strictly negative (see also \cite{HS}), 
so that equality (\ref{GR-average}) is not satisfied and the solution with $R=R_g$ for $r>r_g$ is not consistent.
Then the GR solution can only be an approximate solution, $R\approx R_g$, of Eq. (\ref{trace-approx}) for $r>r_g$.

Applying the divergence theorem to Eq. (\ref{trace-approx}) over the sphere with radius $r_g$, we have
that the integral in Eq. (\ref{GR-average}) is negative if and only if
\begin{equation}\label{negative-deriv-eta-rg}
\frac{d\eta}{dr}(r=r_g) < 0.
\end{equation}
If we now denote by $\eta_g$ the minimizer of the effective potential $V_{\rm eff}(\eta)$ corresponding to $\rho(r)=\rho_g$,
since $\eta$ approximately minimizes $V_{\rm eff}$ in the solar neighbourhood of the galaxy, 
then Eq. (\ref{eta-equation-potential}), using Eq. (\ref{lambda-definition}), becomes
\begin{equation}\label{Yukawa-outskirtsSolS}
\nabla^2\eta \approx \frac{1}{\lambda_g^2}(\eta-\eta_g), \qquad\mbox{for }r>r_g,
\end{equation}
where $\lambda_g = \lambda(\rho_g)$. 
The computations in the present paper will be made under the condition $\lambda_g \gg r_g$ which will permit us
to find analytic estimates of the results.

We require the gravitational field of the Sun to become negligible in comparison with the galactic field at large distances
from the Sun's center, so that we impose the boundary condition
\begin{equation}\label{bound-cond-infinity}
\eta(r) \approx \eta_g, \qquad\mbox{for } \frac{r}{\lambda_g} \gg 1.
\end{equation}
The solution of equation (\ref{Yukawa-outskirtsSolS}) with such a boundary condition
is the Yukawa profile
\begin{equation}
\eta(r) = C \frac{e^{-r/\lambda_g}}{r} + \eta_g, \qquad r>r_g,
\end{equation}
with $C$ constant to be determined by matching the Yukawa profile with the solution for $r<r_g$ that will be computed in the next section.
The constant $C$ measures the deviation of $R$ from $R_g$ in the solar neighborhood of the galaxy. 
Note that if the extremum of the effective potential were a maximum,
then the Yukawa profile would be replaced by a damped (according to $1/r$) trigonometric one.

Using $\lambda_g \gg r_g$ the Yukawa profile is approximated by
\begin{equation}\label{eta-outer-solar-system}
\eta(r) \approx \frac{C}{r} + \eta_g, \qquad r_g<r\ll \lambda_g.
\end{equation}
Note that the condition (\ref{negative-deriv-eta-rg}) imposed by the presence of the Sun requires $C>0$.

{\it Case of specific choice of $f^1,f^2$}.
For the specific choice (\ref{f1-f2-specific}) of functions of curvature
the minimizer $\eta_g$ of the effective potential $V_{\rm eff}(\eta)$, corresponding to $\rho(r)=\rho_g$,
is given by
\begin{equation}\label{eta-g-minimizer}
\eta_g = \frac{c^4}{8\pi G}\left\{ 1-\frac{16\pi G}{c^2}\rho_g\left[ q_1 + \alpha q_2\left(\frac{8\pi G}{c^2}\rho_g\right)^{\alpha-1} \right] \right\}.
\end{equation}
Using Eq. (\ref{lambda-expression}) we see that for $\alpha<1$ the quantity $\lambda(\rho)$ increases as density decreases,
and this is a typical property of the chameleon mechanism \cite{KW}.
Particularly, the Yukawa range $\lambda_g=\lambda(\rho_g)$ is an upper bound for $\lambda$ in the Solar System.

\subsubsection{Solution for $\eta$ in the Sun's interior}\label{sec:eta-sol-interior}

In the region of the Sun's interior where the high-curvature solution approximately holds, according to definition (\ref{eta-definition}),
 $\eta$ is given by
\begin{equation}\label{eta-solution-HC-interior}
\eta(r,R) \approx \eta(r,8\pi G\rho(r)\slash c^2).
\end{equation}
{\it Case of specific choice of $f^1,f^2$}.
For the specific choice (\ref{f1-f2-specific}) of functions of curvature,
using (\ref{eta-definition}) and the expression (\ref{lambda-expression}) of $\lambda^2$, we have
\begin{eqnarray}\label{eta-inner-interior-complete}
\eta &\approx& \frac{c^4}{8\pi G} -2 \left[ q_1 +q_2\alpha\left(\frac{8\pi G}{c^2}\rho\right)^{\alpha-1} \right] \rho c^2 \nonumber\\
&=& \frac{c^4}{8\pi G} -2 \left[ q_1+\frac{\lambda^2(\rho)}{6(1-\alpha)} \right]\rho c^2,
\end{eqnarray}
where
\begin{equation}\label{lambda-rescaled-lambdag}
\lambda^2(\rho) = \lambda_g^2\left(\frac{\rho}{\rho_g}\right)^{\alpha-1}.
\end{equation}
Hence for $\alpha<0$ the length $\lambda(\rho)$ becomes very small in the part of the solar interior where the
high-curvature solution holds, although $\lambda_g\gg r_g$. For instance, if $\alpha=-1$ and $\lambda_g=10^3 r_g \sim 10^5$ AU,
at the bottom of the solar convection zone where $\rho\approx 1.65\times 10^{-1} \,{\rm g}/{\rm cm}^3$, we have
$\lambda(\rho)\approx 10^{-7} \,{\rm m}$. At the top of the convection zone where $\rho\approx 2.73\times 10^{-7} \,{\rm g}/{\rm cm}^3$, we have
$\lambda(\rho)\approx 7\times 10^{-2} \,{\rm m}$, hence completely negligible quantities in both cases (the values of density are computed by using the
density profile in Section \ref{sec:convection} of Appendix B). 

Hence in the case $\alpha<0$ and $|\alpha|$ of order of magnitude unity,
neglecting the contribution of the term proportional to $\lambda^2$, in the high-curvature region we have
\begin{equation}\label{eta-inner-interior}
\eta \approx \frac{c^4}{8\pi G} -2 q_1\rho c^2,
\end{equation}
so that the linear term $q_1 R$ in the NMC function $f^2(R)$ is dominant in the Sun's interior \cite{stellequil}, where density is large enough.
Then, using the trace equation (\ref{trace-approx}), the curvature $R$ is given by
\begin{equation}\label{curvature-inner-interior}
R \approx \frac{8\pi G}{c^2}\,\rho - 48 q_1 \frac{\pi G}{c^2}\,\nabla^2\rho,
\end{equation}
so that, after neglecting $\lambda(\rho)$, the consistency condition (\ref{consist-condition}) becomes
\begin{equation}\label{third-consist-cond}
6 \vert q_1 \vert \, \frac{\vert \nabla^2\rho \vert}{\rho} \ll 1,
\end{equation}
which corresponds to the third condition of Eq. (\ref{consist-decomposed}).
By using the model of mass density profile in the radiative part of the solar interior reported in Section \ref{sec:radiative}
of Appendix B, the maximum of the quantity $|\nabla^2\rho|\slash\rho$ in the radiative interior
is achieved at $r\approx 0.22 R_\odot$ and it is of order of $10^2\slash R_\odot^2$, from which it follows the constraint
\begin{equation}\label{q1-bound-radiat}
\vert q_1 \vert \ll 10^{-2}R_\odot^2 \qquad \mbox{in the radiative interior.}
\end{equation}
In the case of the solar convection zone, using the model of mass density profile reported in Section \ref{sec:convection}
of Appendix B, the maximum of the quantity $|\nabla^2\rho|\slash\rho$ in the convection zone
is achieved at the top of the zone and it is of order of $10^7\slash R_\odot^2$, from which it follows the constraint
\begin{equation}
\vert q_1 \vert \ll 10^{-7}R_\odot^2 \qquad \mbox{in the convection zone.}
\end{equation}
Hence the upper bound on $|q_1|$ in the Sun's interior becomes more stringent as mass density decreases.  

\subsection{Screening mechanism}

In the outer zone of the solar interior (the outer part of the convective zone), in the solar atmosphere, and in the interplanetary space,
the solution $\eta$ interpolates between the value (\ref{eta-solution-HC-interior}) in the inner Sun's interior, and the value (\ref{eta-outer-solar-system})
in the outskirts of the Solar System, respectively.
In order to compute such interpolating function we adapt to NMC gravity the chameleon mechanism developed in Ref. \cite{KW}
and further analysed in Ref. \cite{DS}.

According to Ref. \cite{KW},
in the inner zone of the Sun's interior the solution $\eta(r)$ remains close to the minimizer of the effective potential $V_{\rm eff}$
for $r<r_s$, where $r_s$ is a critical radius, called the screening radius, that has to be determined.
Hence the high-curvature solution holds for $r<r_s$,
GR is approximately satisfied, the consistency condition (\ref{necess-cond-expl}) has also to be satisfied, and
the solution for $\eta$ is given by Eq. (\ref{eta-solution-HC-interior}).

In order to compute $\eta$ for $r>r_s$, first we
integrate $\nabla^2\eta$ over the spherical shell of radii $r_s$ and $r>r_s$, and we use the divergence theorem:
\begin{equation}
4\pi\int_{r_s}^r \nabla^2\eta (r^\prime)^2 dr^\prime =4\pi \left[ \frac{d\eta}{dr}(r)r^2 - \frac{d\eta}{dr}(r_s)r_s^2 \right],
\end{equation}
from which, solving with respect to $d\eta\slash dr$ and integrating, it follows
\begin{eqnarray}
\eta(r) &=& \eta(r_s) + \frac{d\eta}{dr}(r_s)\left( r_s - \frac{r_s^2}{r} \right) \nonumber\\
&+& \int_{r_s}^r \frac{1}{(r^\prime)^2}\int_{r_s}^{r^\prime} \nabla^2\eta(r^{\prime\prime})^2 dr^{\prime\prime}dr^\prime,
\label{eqn:conditionforalphaq1q2}
\end{eqnarray}
then, integrating by parts, using Eq. (\ref{eta-equation-potential}) and the expression of the effective potential $V_{\rm eff}$,
it follows that the function $\eta$ satisfies the integral equation
\begin{widetext}
\begin{equation}\label{eta-integral-equation}
\eta(r) =  \eta(r_s) + \frac{d\eta}{dr}(r_s)\left( r_s - \frac{r_s^2}{r} \right) +
\frac{1}{r}\int_{r_s}^r \left[ \frac{1}{3}\,\rho(r^\prime)c^2-\frac{\partial V}{\partial\eta} \right] \left(r^\prime\right)^2 dr^\prime
- \int_{r_s}^r \left[ \frac{1}{3}\,\rho(r^\prime)c^2-\frac{\partial V}{\partial\eta} \right] r^\prime dr^\prime,
\end{equation}
\end{widetext}
where $\eta(r_s)$ and $d\eta\slash dr(r_s)$ are evaluated by using Eq. (\ref{eta-solution-HC-interior}).

Now we proceed to estimate the expression in square brackets in the integrands.
For $r>r_s$, in the outer zone of the Sun's interior and in the inner solar atmosphere, 
where mass density is significantly larger than the galactic density $\rho_g$, 
we require the potential $V(\eta,\rho)$ to satisfy the condition (see also Ref. \cite{KW})
\begin{equation}\label{deriv-potential-condit}
\left\vert \frac{\partial V}{\partial\eta}(\eta,\rho) \right\vert \ll \frac{1}{3}\,\rho c^2,
\end{equation} 
as soon as $\eta$ is displaced enough from the minimizer of $V_{\rm eff}$.
The explicit dependence of $\partial V/\partial\eta$ on density $\rho$ is a distinctive feature of the
application of the chameleon mechanism to NMC gravity with respect to $f(R)$ gravity.

Using Eq. (\ref{eta-equation-potential}), condition (\ref{deriv-potential-condit}) reads
\begin{equation}\label{omega-condition}
\frac{c^2}{8\pi G}\,\omega(\eta,\rho) \ll \rho.
\end{equation}
This inequality will be verified a posteriori in the case of the specific choice (\ref{f1-f2-specific}) of functions of curvature.
Given that the proof is a bit technical, it is given in Section \ref{sec:ineq-V-rho} of Appendix A.

Since the function $\omega(\eta,\rho)$ is equal to curvature $R$, then $R$ is small in comparison with the GR solution
and we say that Eq. (\ref{omega-condition}) is the condition
for a local low-curvature solution as in Ref. \cite{HS}.
Hence, for $r>r_s$ the solution locally deviates from GR, while deviations from GR are {\it screened} for $r<r_s$.
We also say that the Sun is screened for $r<r_s$.

In the outer solar atmosphere and interplanetary space, where mass density becomes smaller and gradually approaches
the galactic density $\rho_g$, we expand the derivative of the potential around the minimizer $\eta_g$:
\begin{equation}\label{potential-deriv-approx}
\frac{\partial V}{\partial\eta}(\eta,\rho) \approx \frac{\partial V}{\partial\eta}(\eta_g,\rho) +
\frac{\partial^2 V}{\partial\eta^2}(\eta_g,\rho)(\eta-\eta_g).
\end{equation}

{\it Case of specific choice of $f^1,f^2$}.
For the specific choice (\ref{f1-f2-specific}) of functions of curvature, the function $\omega(\eta,\rho)$, that has to be used in condition
(\ref{omega-condition}), is given by
\begin{eqnarray}\label{R-omega-formula}
\omega(\eta,\rho) &=& \left( \frac{16\pi G}{c^2} \, \alpha q_2\rho \right)^{1\slash(1-\alpha)} \nonumber\\
&\times& \left(1-\frac{16\pi G}{c^2}\,q_1\rho - \frac{8\pi G}{c^4}\,\eta \right)^{1\slash(\alpha-1)}.
\end{eqnarray}

\subsubsection{Solution for $\eta$ in the case of specific functions $f^1,f^2$}

In the sequel we consider the specific choice (\ref{f1-f2-specific}) of functions of curvature.
We assume $\alpha<0$ and $|\alpha|$ of order of magnitude unity, moreover, 
we require the consistency condition (\ref{consist-condition}) to be satisfied for $r<r_s$ in the Sun's interior.

Using now (\ref{f1-f2-specific}) and the definition (\ref{eta-definition}) of $\eta$ we have
\begin{equation}\label{eta-spec-expression}
\eta = \frac{c^4}{8\pi G} - 2\left( q_1 + \alpha q_2 R^{\alpha-1} \right) \rho c^2.
\end{equation}
In Section \ref{sec:eta-sol-interior} we have argued that the linear term $q_1 R$ in the NMC function $f^2(R)$ is dominant in the Sun's interior,
where density is large enough.
Conversely, in regions with low mass density, such as galactic and interplanetary space and even the solar atmosphere,
the effect of the term $q_2 R^\alpha$ becomes dominant for $\alpha<0$ as it will be shown in the sequel 
(see also Ref. \cite{stellequil} and the discussion in Ref. \cite{curraccel}).
Then, because of the smallness of the spacetime curvature $R$ in the outer solar atmosphere and interplanetary space,
we assume
\begin{equation}\label{q1-ll-bound}
|q_1| \ll |q_2| R^{\alpha-1},
\end{equation}
so that in this region of space $\eta$ is given by
\begin{equation}\label{eta-approx-interpl-space}
\eta \approx \frac{c^4}{8\pi G} - 2 \alpha q_2 R^{\alpha-1} \rho c^2.
\end{equation}
The validity of assumption (\ref{q1-ll-bound}) will be verified a posteriori by resorting to the constraint from Cassini measurement.
Since the proof is a bit technical it is given in Section \ref{sec:q1-q2-Rexp-alpha} of Appendix A.
Solving Eq. (\ref{eta-approx-interpl-space}) with respect to curvature $R=\omega(\eta,\rho)$ we find
\begin{equation}
\omega(\eta,\rho) \approx \left( \frac{16\pi G}{c^2} \alpha q_2\rho \right)^{1\slash (1-\alpha)}
\left( 1 - \frac{8\pi G}{c^4} \eta \right)^{1\slash(\alpha-1)},
\end{equation}
from which, using (\ref{eta-equation-potential}) we obtain the property
\begin{equation}
\frac{\partial V}{\partial\eta}(\eta,\rho) \approx \left( \frac{\rho}{\rho_g} \right)^{1\slash(1-\alpha)} \frac{\partial V}{\partial\eta}(\eta,\rho_g).
\end{equation}
Taking now into account that at density $\rho_g$ (in the solar vicinity of the galaxy) the field $\eta$ approximately
minimizes the effective potential $V_{\rm eff}$, so that
\begin{equation}
\frac{\partial V}{\partial\eta}(\eta_g,\rho_g) \approx \frac{1}{3}\, \rho_g c^2,
\end{equation}
we can compute the approximation (\ref{potential-deriv-approx}) of the derivative of the potential:
\begin{widetext}
\begin{eqnarray}\label{potential-deriv-computed}
\frac{\partial V}{\partial\eta}(\eta,\rho) &\approx& \left( \frac{\rho}{\rho_g} \right)^{1\slash(1-\alpha)}
\left[ \frac{\partial V}{\partial\eta}(\eta_g,\rho_g) +
\frac{\partial^2 V}{\partial\eta^2}(\eta_g,\rho_g)(\eta-\eta_g) \right]
\approx \left( \frac{\rho}{\rho_g} \right)^{1\slash(1-\alpha)} \left[ \frac{1}{3}\, \rho_g c^2 + \frac{1}{\lambda_g^2}(\eta-\eta_g) \right] \nonumber\\
&\approx& \frac{1}{3}\, \rho_g c^2 \left( \frac{\rho}{\rho_g} \right)^{1\slash(1-\alpha)},
\end{eqnarray}
\end{widetext}
where we have taken into account that $\lambda_g$ is assumed large.

We now proceed to solve the equation for $\eta$ by evaluating the integrals in Eq. (\ref{eta-integral-equation}). 
For $r>r_s$, in the outer zone of the Sun's interior and in the inner solar atmosphere, using Eq. (\ref{deriv-potential-condit}) we have
\begin{equation}\label{small-V-derivative}
\frac{1}{3}\, \rho c^2 - \frac{\partial V}{\partial\eta} \approx \frac{1}{3}\, \rho c^2.
\end{equation}
In the outer solar atmosphere and interplanetary space, using (\ref{potential-deriv-computed}) we have
\begin{equation}\label{sq-bra-NMC}
\frac{1}{3}\, \rho c^2 - \frac{\partial V}{\partial\eta} \approx
\frac{1}{3}\, \rho c^2 \left[ 1 - \left(\frac{\rho_g}{\rho}\right)^{-\alpha\slash(1-\alpha)} \right].
\end{equation}
Since we have
\begin{equation}
\alpha < 0 \implies 0 < -\frac{\alpha}{1-\alpha} < 1,
\end{equation}
for $|\alpha|$ not too small, the second term inside the square bracket in Eq. (\ref{sq-bra-NMC}) is negligible in comparison to 1
in the outer zone of the Sun's interior and in the inner solar atmosphere,
where mass density is significantly larger than the galactic density $\rho_g$,
so that Eq. (\ref{sq-bra-NMC}) is valid with a good approximation for any $r>r_s$.

We are ready to write the solution for $\eta$ in terms of the screening radius.
Using Eq. (\ref{eta-inner-interior-complete}), for $r<r_s$ we have the solution
\begin{equation}\label{eta-solution-below-rs}
\eta =  \frac{c^4}{8\pi G} -2 \left[ q_1+\frac{\lambda^2(\rho)}{6(1-\alpha)} \right]\rho c^2,
\end{equation}
which satisfies the boundary condition (\ref{bound-cond-0}) by using the expression (\ref{lambda-expression}) of $\lambda^2(\rho)$
and the density model of the Sun's radiative interior in Section \ref{sec:radiative} of Appendix B, which yields $d\rho\slash dr = 0$ at $r=0$.

Then we introduce the effective mass $M_{\rm eff}(r)$ defined for $r>r_s$ as follows:
\begin{equation}\label{effective-mass}
M_{\rm eff}(r) = 4\pi \int_{r_s}^r \rho \left[ 1 - \left(\frac{\rho_g}{\rho}\right)^{-\alpha\slash(1-\alpha)} \right] (r^\prime)^2 dr^\prime.
\end{equation}

Substituting now the expression (\ref{sq-bra-NMC}) in Eq. (\ref{eta-integral-equation}), and using Eqs. (\ref{eta-solution-below-rs}-\ref{effective-mass}),
we obtain the solution $\eta$ for $r_s<r<r_g$:
\begin{widetext}
\begin{eqnarray}\label{eta-solution-complete}
\eta(r) &=& \frac{c^4}{8\pi G} -2c^2 \left\{ \left[ q_1+\frac{\lambda_s^2}{6(1-\alpha)} \right]\rho_s + 
\left[ q_1+\frac{\alpha\lambda_s^2}{6(1-\alpha)} \right] \rho_s^\prime\left( r_s - \frac{r_s^2}{r} \right) \right\}
+ \frac{c^2}{12\pi} \, \frac{M_{\rm eff}(r)}{r} \nonumber\\
&-& \frac{c^2}{3}\int_{r_s}^r \rho \left[ 1 - \left(\frac{\rho_g}{\rho}\right)^{-\alpha\slash(1-\alpha)} \right] r^\prime dr^\prime,
\end{eqnarray}
\end{widetext}
where $\rho_s=\rho(r_s)$, $\rho_s^\prime=d\rho\slash dr(r_s)$ and $\lambda_s=\lambda(\rho_s)$.
Here we have kept the terms involving $\lambda_s$, notwithstanding that $\lambda_s$ is negligibly small (see Section \ref{sec:eta-sol-interior}), 
because such terms will be necessary for the verification a posteriori of inequality (\ref{omega-condition}),
which is a crucial property of a chameleon solution \cite{KW}. Since this is the only reason to keep the terms
with $\lambda_s$ in the expression of $\eta$, in the following formulae we neglect such terms.

The solution (\ref{eta-solution-below-rs}) and (\ref{eta-solution-complete}) is continuous with its derivative at $r=r_s$, 
moreover, $\eta$ has to be continuous with its derivative at $r=r_g$.
Imposing the continuity of the derivative in $r_g$, and using the expression (\ref{eta-outer-solar-system}) of $\eta$ for $r>r_g$,
the constant $C$ in Eq. (\ref{eta-outer-solar-system}) is determined:
\begin{equation}\label{formula-for-C}
C = \frac{c^2}{12\pi} \, M_{\rm eff}(r_g) + 2 c^2 q_1\rho_s^\prime r_s^2.
\end{equation}

Imposing the continuity of $\eta$ in $r_g$ the integral identity follows:
\begin{eqnarray}\label{integral-identity-r-screen}
\frac{c^2}{8\pi G} &-& 2q_1(\rho_s + \rho_s^\prime r_s) - \frac{\eta_g}{c^2} \nonumber\\
&=& \frac{1}{3}\int_{r_s}^{r_g} \rho \left[ 1 - \left(\frac{\rho_g}{\rho}\right)^{-\alpha\slash(1-\alpha)} \right] r dr.
\end{eqnarray}
The solution $\eta$ is completely determined once the screening radius $r_s$ is determined. Then,
using the expression (\ref{eta-g-minimizer}) of the minimizer $\eta_g$ we have
\begin{equation}\label{eta-g-rewritten}
\frac{\eta_g}{c^2} = \frac{c^2}{8\pi G} -2q_1\rho_g - 2\alpha q_2\left(\frac{8\pi G}{c^2}\right)^{\alpha-1} \rho_g^\alpha.
\end{equation}
Substituting in Eq. (\ref{integral-identity-r-screen}), and taking into account that $\rho_g \ll \rho(r_s)$, 
we find an integral equation which determines the screening radius $r_s$ and completes the solution for $\eta$:
\begin{eqnarray}\label{rs-integr-equation}
& &\frac{1}{6}\int_{r_s}^{r_g} \rho \left[ 1 - \left(\frac{\rho_g}{\rho}\right)^{-\alpha\slash(1-\alpha)} \right] r dr \\
&=& -q_1(\rho_s + \rho_s^\prime r_s) + \alpha q_2 \left(\frac{8\pi G}{c^2}\right)^{\alpha-1}\rho_g^\alpha. \nonumber
\end{eqnarray}
This is the NMC version of the integral equation found in Ref. \cite{DS} for the chameleon mechanism.
Eventually, the solution for $\eta$ is given by formula (\ref{eta-solution-complete}) for $r_s<r<r_g$.

\subsubsection{Verification of inequalities}

The solution for $\eta$ has been computed by assuming some inequalities, necessary in order to find an analytic
approximation of the solution, that have to be verified a posteriori. Thus, we now show that the computed solution
satisfies some of these inequalities (the inequalities that require a technical proof are verified in Appendix A).

We see immediately that the expressions (\ref{eta-solution-below-rs}) and (\ref{eta-solution-complete}),
which give the solution for $r\leq r_g$, satisfy
\begin{equation}\label{eta-O-property}
\left\vert \frac{8\pi G}{c^4}\eta -1 \right\vert = \OO\left(\frac{1}{c^2}\right) \ll 1,
\end{equation}
which yields inequality (\ref{cond-f1R-f2R}).
Moreover, using formula (\ref{formula-for-C}) for the constant $C$, and formula (\ref{eta-g-minimizer}) for the minimizer $\eta_g$
together with the integral equation (\ref{rs-integr-equation}), we see that
also the solution (\ref{eta-outer-solar-system}) satisfies the same inequality for $r>r_g$.
Hence the computed solution $\eta$ satisfies inequality (\ref{cond-f1R-f2R}) for any $r$.

Using now the expression (\ref{curvature-inner-interior}) of curvature, for $r<r_s$ we have $R = \OO(1\slash c^2)$.
Then, the integral equation (\ref{rs-integr-equation}) implies
\begin{equation}\label{q2-O-property}
\vert q_2 \vert = \rho_g^{-\alpha} \cdot \OO\left(\frac{1}{c^{2-2\alpha}}\right) \ll 1,
\end{equation}
being $\alpha<0$, from which, using formula (\ref{R-omega-formula}) for curvature $R=\omega(\eta,\rho)$
and property (\ref{eta-O-property}) for $\eta$, we have $R = \OO(1\slash c^2)$ for $r_s<r<r_g$.
Eventually, since $R\approx R_g$ for $r>r_g$ this property is satisfied for any $r$.

Now we observe that the quantity $|q_1|R$ is largest at the Sun's centre and, using inequality (\ref{q1-bound-radiat}),
we have
\begin{equation}
\vert q_1 \vert R(0) \ll 10^{-2}R_\odot^2 \frac{8\pi G}{c^2}\,\rho(0) \ll 10^{-5},
\end{equation}
so that, from definition (\ref{f1-f2-specific}) of $f^1$ and $f^2$, using property (\ref{q2-O-property}) it follows
\begin{equation}
\left\vert f^2(R) \right\vert = \left\vert q_1R + q_2R^\alpha \right\vert < \vert q_1 \vert R(0) + \rho_g^{-\alpha} \cdot \OO\left(\frac{1}{c^2}\right) \ll 1,
\end{equation}
which yields the second of inequalities (\ref{cond-f1-f2}), the first being trivial.

\subsubsection{Solution for the potentials $\Phi$ and $\Psi$}

We have assumed that the high-curvature solution holds for the gravitational field of the galaxy in the
solar neighbourhood for $r>r_g$, so that GR is approximately satisfied. 
This assumption implies that the Milky Way is screened within a distance
of about $8\,{\rm kpc}$ from its center, where the Solar System is approximately located. Such a screening condition
may impose additional constraints on the NMC gravity model whose assessment requires the solution for
the gravitational field of the Milky Way, possibly taking also into account the effect of the other
galaxies in the local group, however that will be the object of future research.

In what follows we denote by $U$ the Newtonian potential of the mass distribution with density $\rho$,
\begin{equation}\label{Newt-potential}
U = G \int \frac{\rho(\mathbf{y})}{|\mathbf{x}-\mathbf{y}|}d^3y,
\end{equation}
which satisfies the Poisson equation $\nabla^2U = - 4\pi G \rho$.

Using Eqs. (\ref{Phi-equation-approx}-\ref{Psi-equation-approx}) it follows that the potential $\Psi$ of the metric is related to the deviation from GR, then
we impose the following boundary conditions in the galaxy at large distances from the Sun's center, where GR is satisfied by our assumptions:
\begin{equation}\label{Phi-Psi-boundcond-infty}
\Phi(r) \approx \frac{1}{c^2}U(r), \quad \Psi(r) \approx 0, \qquad\mbox{for } \frac{r}{\lambda_g} \gg 1.
\end{equation}

Combining equations (\ref{Psi-equation-approx}) and (\ref{trace-approx}) for $\Psi$ and $\eta$ we have
\begin{equation}
\nabla^2 \left( \Psi + \frac{8\pi G}{c^4}\, \eta \right) = 0,
\end{equation}
which implies in the case of spherical symmetry:
\begin{equation}
\Psi + \frac{8\pi G}{c^4}\, \eta = a_1 + \frac{a_2}{r},
\end{equation}
with $a_1,a_2$ constants to be determined. Requiring $\Psi$ to be not singular at $r=0$ imposes $a_2=0$.

Using the boundary conditions (\ref{bound-cond-infinity}) and (\ref{Phi-Psi-boundcond-infty}) for $\eta$ and $\Psi$, respectively, it then follows
\begin{equation}
a_1 = \frac{8\pi G}{c^4}\, \eta_g,
\end{equation}
from which we find
\begin{equation}
\Psi(r) = - \frac{8\pi G}{c^4}\left[ \eta(r) - \eta_g \right].
\end{equation}
The solution for $\Psi$ then follows immediately from the solution for $\eta$: 
using Eqs. (\ref{eta-solution-below-rs}), (\ref{eta-g-rewritten}) and (\ref{q1-ll-bound}) with $R=R_g$, we find for $r<r_s$,
\begin{equation}\label{Psi-solution-interior}
\Psi(r) = 16\frac{\pi G}{c^2} \left[ q_1\rho(r) - \alpha q_2 R_g^{\alpha-1}\rho_g \right],
\end{equation}
using Eqs. (\ref{eta-solution-complete}) and (\ref{eta-g-rewritten}-\ref{rs-integr-equation}), and neglecting $\lambda_s$, we find for $r_s<r<r_g$,
\begin{widetext}
\begin{equation}\label{Psi-solution-solar-sys}
\Psi(r) = -16 \frac{\pi G}{c^2} q_1\rho_s^\prime\frac{r_s^2}{r} -\frac{2G}{3c^2}\,\frac{M_{\rm eff}(r)}{r}
- \frac{8\pi G}{3c^2}\int_r^{r_g} \rho \left[ 1 - \left(\frac{\rho_g}{\rho}\right)^{-\alpha\slash(1-\alpha)} \right] r^\prime dr^\prime,
\end{equation}
\end{widetext}
and, using Eq. (\ref{eta-outer-solar-system}) and formula (\ref{formula-for-C}), we find for $r>r_g$,
\begin{equation}\label{Psi-outer-solution}
\Psi(r) = -\frac{2G}{3c^2}\,\frac{M_{\rm eff}(r_g)}{r} - 16\frac{\pi G}{c^2}q_1\rho_s^\prime\frac{r_s^2}{r}.
\end{equation}
Combining now equations (\ref{Phi-equation-approx}) and (\ref{Psi-equation-approx}) for $\Phi$ and $\Psi$ we have
\begin{equation}
\nabla^2\left( \Phi - \frac{U}{c^2} -\frac{1}{2}\, \Psi \right) = 0.
\end{equation}
The solution $\Phi$ of this equation which is not singular at $r=0$ and satisfies the boundary conditions (\ref{Phi-Psi-boundcond-infty})
is given by
\begin{equation}\label{Phi-solution-everywhere}
\Phi = \frac{U}{c^2} + \frac{1}{2}\, \Psi.
\end{equation}
The solutions found for $\Phi$ and $\Psi$ define the space-time metric (\ref{metric}).

\subsubsection{Screening radius in the convection zone}\label{sec:screen-convect}

In this section we consider the case of the screening radius in the solar convection zone which is relevant for the
constraint from the Cassini measurement. Using the results of Section \ref{sec:eta-sol-interior}, we neglect $\lambda$
in the convection zone so that the consistency condition (\ref{consist-condition}) becomes inequality (\ref{third-consist-cond})
which we write in the form
\begin{equation}\label{consist-cond-10-3}
\vert q_1 \vert \, \frac{\vert \nabla^2\rho \vert}{\rho} < \varepsilon \ll 1.
\end{equation}
For $r>R_\odot$ we consider the term
\begin{equation}\label{sign-interest}
-\frac{2G}{rc^2}\left[ 8\pi q_1\rho_s^\prime r_s^2 +\frac{1}{3}\,M_{\rm eff}(r) \right]
\end{equation}
in the expression (\ref{Psi-solution-solar-sys}) of $\Psi(r)$. 
We approximate the effective mass (\ref{effective-mass}) for $r>R_\odot$, and the integral in Eq. (\ref{rs-integr-equation}), with
\begin{eqnarray}
& &M_{\rm eff}(r) \approx M_{\rm eff}(R_\odot) \approx 4\pi\int_{r_s}^{R_\odot} \rho(r) r^2 dr, \label{first-integral}
\\
& &\int_{r_s}^{r_g} \rho \left[ 1 - \left(\frac{\rho_g}{\rho}\right)^{-\alpha\slash(1-\alpha)} \right] r dr \approx
\int_{r_s}^{R_\odot}\rho(r)rdr, \label{second-integral}
\end{eqnarray}
where we have neglected the contribution from the solar atmosphere outside the photosphere, and
we have taken into account that $\rho_g \ll \rho(r)$ for $r\leq R_\odot$ and $\alpha<0$.

Then, using inequality (\ref{consist-cond-10-3})  and the expressions of $\rho$ and $\nabla^2\rho$ in the convection zone 
reported in Section \ref{sec:convection} of Appendix B, we find that
\begin{eqnarray}
8 \pi \left\vert q_1\rho_s^\prime \right\vert r_s^2 &<& \frac{20}{3} \varepsilon M_{\rm eff}(R_\odot), \label{zconv-first-ineq-eps}
\\
|q_1|\cdot |\rho_s+\rho_s^\prime r_s| &<& 20\varepsilon \, \frac {1}{6}\int_{r_s}^{R_\odot}\rho(r)r dr, \label{zconv-second-ineq-eps}
\end{eqnarray}
for any $r_s$ in the convection zone. Hence, for $\varepsilon\leq 1/20$
the quantity (\ref{sign-interest}) is negative and, using (\ref{Psi-solution-solar-sys}-\ref{Psi-outer-solution}), it follows $\Psi(r)<0$ for any $r>R_\odot$.

Moreover, for such values of $\varepsilon$, using Eq. (\ref{formula-for-C}), it follows that the constant $C$ in Eq. (\ref{eta-outer-solar-system}) is positive, in agreement with
the statement immediately following Eq. (\ref{eta-outer-solar-system}).
Hence, it follows that condition (\ref{negative-deriv-eta-rg}) is satisfied, so that the integral in Eq. (\ref{GR-average})
is negative as anticipated in Section \ref{Sec:outskirts}.

Eventually, with the above approximations the solution (\ref{eta-solution-complete}) for $\eta$ in the interval $(R_\odot,r_g)$
takes a simpler form, which, also neglecting $\lambda_s$, we write as
\begin{eqnarray}\label{eta-convz-simple}
1-\frac{8\pi G}{c^4}\eta &\approx& 2\alpha q_2 \left(\frac{8\pi G}{c^2}\rho_g\right)^\alpha - \frac{16\pi G}{c^2} q_1\rho_s^\prime\frac{r_s^2}{r} \nonumber\\
&-& \frac{2G}{3c^2}\,\frac{M_{\rm eff}(R_\odot)}{r}.
\end{eqnarray}
This approximation will be used in the sequel so to compute in the solar atmosphere the extra force due to the nonminimal coupling.

\section{Fifth force and extra force}

In this section we compute the equations of hydrodynamics of a perfect fluid by resorting to the covariant divergence
of the energy-momentum tensor \cite{BBHL}, as given by Eq. (\ref{covar-div-1}) that we repeat for convenience:
\begin{equation}\label{covar-div-2}
\nabla_\mu T^{\mu\nu} = \frac{f^2_R }{ 1 + f^2} ( g^{\mu\nu} \LL_m - T^{\mu\nu} ) \nabla_\mu R.
\end{equation}
First we compute the $\nu=t$ component of this equation. Using the components of the energy-momentum tensor
given by Eqs. (\ref{T-tt-tr})-(\ref{T-other-mixed}), the left-hand side of Eq. (\ref{covar-div-2}) yields
\begin{equation}
\frac{1}{c}\,\nabla_\mu T^{\mu t} =
\frac{\partial\rho}{\partial t}+\frac{\partial}{\partial r}(\rho v) +\frac{2}{r}\,\rho v + \OO\left(\frac{1}{c^2}\right).
\end{equation}
In order to compute the right-hand side of Eq. (\ref{covar-div-2}), first we observe that, using definition (\ref{f1-f2-specific}) of $f^2$
and property (\ref{q2-O-property}) of $q_2$, we have
\begin{equation}
f^2_R = q_1 + \alpha q_2 R^{\alpha-1} = q_1 + \rho_g^{-\alpha} \cdot \OO(1) = \OO(1),
\end{equation}
with respect to $1/c^2$. Then, taking into account that $f^2$ satisfies the second of inequalities (\ref{cond-f1-f2}),
the evaluation of the right-hand side of Eq. (\ref{covar-div-2}) yields
\begin{equation}
\frac{f^2_R }{ 1 + f^2} ( g^{\mu t} \LL_m - T^{\mu t} ) \frac{\partial R}{\partial x^\mu} = \OO\left(\frac{1}{c}\right).
\end{equation}
Neglecting terms of order $\OO(1\slash c^2)$ the continuity equation, in the case of spherical symmetry and radial motion,
then in the nonrelativistic limit, it follows as usual:
\begin{equation}\label{contin-eq}
\frac{\partial\rho}{\partial t}+\frac{\partial}{\partial r}(\rho v) +\frac{2}{r}\,\rho v = 0.
\end{equation}
The NMC term on the right-hand side of Eq. (\ref{covar-div-2}) gives a distinctive contribution
to the spatial part of this equation that now we compute. For $\nu=r$, using the components (\ref{T-tt-tr})-(\ref{T-other-mixed})
of the energy-momentum tensor, the left-hand side yields
\begin{eqnarray}
\nabla_\mu T^{\mu r} &=& \frac{\partial}{\partial t}(\rho v) + \frac{\partial}{\partial r}(\rho v^2)
-\rho c^2 \left( \frac{d\Phi}{dr} - \frac{d\Psi}{dr} \right) \nonumber\\
&+& \frac{2}{r}\,\rho v^2 + \frac{\partial p}{\partial r} + \OO\left(\frac{1}{c^2}\right).
\end{eqnarray}
Using now the continuity equation Eq. (\ref{contin-eq}) at order $\OO(1)$, and the solution (\ref{Phi-solution-everywhere}) for the metric potential $\Phi$, 
we get
\begin{equation}\label{space-covar-div}
\nabla_\mu T^{\mu r} = \rho\frac{dv}{dt} + \frac{\partial p}{\partial r} - \rho\frac{dU}{dr} + \frac{1}{2} \rho c^2 \frac{d\Psi}{dr}
+ \OO\left(\frac{1}{c^2}\right),
\end{equation}
where $d/dt=\partial/\partial t+v\partial/\partial r$ is the time derivative {\it following the fluid}.

The right-hand side of Eq. (\ref{covar-div-2}) yields
\begin{equation}\label{extra-f-r}
\frac{f^2_R }{ 1 + f^2} ( g^{\mu r} \LL_m - T^{\mu r} ) \frac{\partial R}{\partial x^\mu} = -c^2 f^2_R\,\rho\frac{dR}{dr} + \OO\left(\frac{1}{c^2}\right).
\end{equation}
Eventually, the components $\nu=\theta$ and $\nu=\varphi$ yield terms of order $\OO(1\slash c^2)$.

Combining equations (\ref{space-covar-div}) and (\ref{extra-f-r}), and neglecting terms of order $\OO(1\slash c^2)$,
we obtain the radial equation of NMC hydrodynamics for a perfect fluid in the nonrelativistic limit, which in the stationary case reads
\begin{equation}\label{NMC-hydro}
\rho\frac{dv}{dt} = \rho\frac{dU}{dr} - \frac{dp}{dr} - \frac{1}{2} \rho c^2 \frac{d\Psi}{dr} - c^2 f^2_R\,\rho\frac{dR}{dr}.
\end{equation}
We observe the presence of two additional terms in comparison with Eulerian equations of Newtonian hydrodynamics:
\begin{itemize}
\item[{\rm (i)}] a fifth force density proportional to the gradient of the metric potential $\Psi$;
\item[{\rm (ii)}] an extra force density proportional to the product of $f^2_R$ by the gradient of curvature $R$.
\end{itemize}
The extra force density in (ii) has been extensively discussed in Ref. \cite{BBHL},
and for relativistic perfect fluids in Ref. \cite{BLP}.
While the fifth force is typical of $f(R)$ gravity theory, the extra force is specific of NMC gravity.
Now we derive the explicit expressions of such force densities corresponding to
the specific choice (\ref{f1-f2-specific}) of functions of curvature.

\subsection{Forces inside the screening radius}

If $r<r_s$, using the solution (\ref{Psi-solution-interior}) for $\Psi$, we find for the fifth force:
\begin{equation}
F_{\rm f} = -8\pi G q_1 \rho \frac{d\rho}{dr}.
\end{equation}
In order to compute the extra force we need $f^2_R$ which,
using the consistency condition (\ref{third-consist-cond}), is approximated by means of the GR value of curvature:
\begin{equation}
f^2_R \approx q_1 + \alpha q_2 \left(\frac{8\pi G}{c^2}\rho\right)^{\alpha-1} = q_1 + \frac{\lambda^2(\rho)}{6(1-\alpha)} \approx q_1,
\end{equation}
being $\lambda^2(\rho)$ negligible inside the screening radius.
Using the solution (\ref{curvature-inner-interior}) for the curvature inside the screening radius, we have for the derivative:
\begin{equation}
\frac{dR}{dr} = \frac{8\pi G}{c^2}\frac{d\rho}{dr} - 48 q_1 \frac{\pi G}{c^2}\frac{d\nabla^2\rho}{dr},
\end{equation}
and neglecting the term with $q_1^2$ (see also Ref. \cite{stellequil} for analogous computations), we find for the extra force
\begin{equation}
F_{\rm e} = -8\pi G q_1 \rho\frac{d\rho}{dr},
\end{equation}
which shows that the fifth force and the extra force are equal inside the screening radius.

\subsection{Forces outside the screening radius}

If $r_s<r<r_g$, using the solution (\ref{Psi-solution-solar-sys}) for $\Psi$, we find for the fifth force:
\begin{equation}\label{Fifth-force-outside-rs}
F_{\rm f} = -8\pi G q_1 \rho_s^\prime \rho(r) \left(\frac{r_s}{r}\right)^2 - \frac{G}{3}\rho(r)\frac{M_{\rm eff}(r)}{r^2}.
\end{equation}
The extra force for $r_s<r<r_g$ is obtained by computing the derivative of the curvature $R=\omega(\eta,\rho)$, given by formula  
(\ref{R-omega-formula}), and using the solution (\ref{eta-solution-complete}) for $\eta$ where we neglect the terms with $\lambda_s$:
\begin{widetext}
\begin{eqnarray}
F_{\rm e} &=& -\frac{c^4}{(1-\alpha)16\pi G}\left(\frac{16\pi G}{c^2}\alpha q_2\right)^{\frac{1}{1-\alpha}}\left(1-\frac{8\pi G}{c^4}\eta\right)
\left( 1-\frac{16\pi G}{c^2}q_1\rho-\frac{8\pi G}{c^4}\eta \right)^{\frac{\alpha-2}{1-\alpha}} \nonumber\\
&\times& \left[ \frac{d\rho}{dr}\left(1-\frac{8\pi G}{c^4}\eta\right) - \frac{16\pi G}{c^2}q_1\rho_s^\prime\rho\left(\frac{r_s}{r}\right)^2 -
\frac{2G}{3c^2}\rho\frac{M_{\rm eff}(r)}{r^2} \right] \rho^{\frac{\alpha}{1-\alpha}}.
\end{eqnarray}
\end{widetext}
In the case of screening radius in the convection zone,
using the approximation (\ref{eta-convz-simple}) of the solution for $\eta$, we can further approximate the term between square brackets
in the expression of $F_{\rm e}$ as follows:
\begin{eqnarray}
&-& \frac{2G}{r c^2}\left( 8\pi q_1\rho_s^\prime r_s^2 + \frac{1}{3}M_{\rm eff}(R_\odot) \right)
\left(\frac{d\rho}{dr}+\frac{\rho}{r}\right) \nonumber\\
&+&2\alpha q_2 R_g^\alpha \frac{d\rho}{dr}.
\end{eqnarray}
Let us now consider $r$ varying in the solar atmosphere, particularly in regions close to the Sun's edge where the density gradient is large.
Using the model of mass density profile in the chromosphere and in the chromosphere-corona transition region,
reported in Section \ref{sec:chromosph} and Section \ref{sec:chromo-coro-transr} of Appendix B, respectively, we observe that
in such regions the density gradient is large enough to have
\begin{equation}
\left\vert\frac{d\rho}{dr}\right\vert \gg \frac{\rho}{r}.
\end{equation}
Consequently, the extra force is approximated by
\begin{eqnarray}
F_{\rm e} &\approx& -\frac{c^4}{(1-\alpha)16\pi G}\left(\frac{16\pi G}{c^2}\alpha q_2\right)^{\frac{1}{1-\alpha}}\left(1-\frac{8\pi G}{c^4}\eta\right)^2 \nonumber\\
&\times& \left( 1-\frac{16\pi G}{c^2}q_1\rho-\frac{8\pi G}{c^4}\eta \right)^{\frac{\alpha-2}{1-\alpha}} \rho^{\frac{\alpha}{1-\alpha}} \,\frac{d\rho}{dr},
\end{eqnarray}
where the expression (\ref{eta-convz-simple}) has to be used for function $\eta$.

In Section \ref{sec:q1-q2-Rexp-alpha} of Appendix A, values of $r_s$ in the convection zone, $r_0$ in the inner region of solar atmosphere, 
and $\alpha$, are given such that the term with $q_1$ can be neglected for any $r>r_0$:
\begin{equation}
\frac{16\pi G}{c^2} |q_1|\rho(r) \ll 1-\frac{8\pi G}{c^4}\eta(r).
\end{equation}
Now we observe that for $r$ in the solar atmosphere and close enough to the Sun's edge the quantity $1\slash r$ is slowly varying in comparison with
density $\rho(r)$ and its derivative, so that we may approximate the expression (\ref{eta-convz-simple}) of $\eta(r)$ with $\eta(R_\odot)$.
Then we obtain the further approximation of the extra force:
\begin{equation}\label{extra-force-outside-rs}
F_{\rm e} \approx \mathcal{F}(\alpha,q_1,q_2) \rho^{\frac{\alpha}{1-\alpha}} \,\frac{d\rho}{dr},
\end{equation}
where the coefficient
\begin{eqnarray}\label{cal-F-eforce-approxim}
\mathcal{F}(\alpha,q_1,q_2) &=& -\frac{c^4}{(1-\alpha)16\pi G}\left(\frac{16\pi G}{c^2}\alpha q_2\right)^{\frac{1}{1-\alpha}} \nonumber\\
&\times& \left[1-\frac{8\pi G}{c^4}\eta(R_\odot)\right]^{\frac{\alpha}{\alpha-1}}
\end{eqnarray}
is independent of $r$. This last approximation of the extra force can be used in regions of the solar atmosphere with high density gradient,
such as the chromosphere and the chromosphere-corona transition region.

\section{Constraint from Cassini measurement}\label{sec:Cassini-measure}

In this section we assume the screening radius in the convection zone and $\varepsilon$ in inequality (\ref{consist-cond-10-3}) such that
$\varepsilon\leq 1/20$ so that $\Psi(r)<0$ for $r>R_\odot$.
Indeed, we will find that the screening radius that saturates the Cassini constraint lies in the convection zone.
Using the metric (\ref{metric}) we have for PPN parameter $\gamma$:
\begin{equation}\label{gamma-expression}
1-\gamma = - \frac{\Psi}{\Phi-\Psi}.
\end{equation}
Since $\Psi(r)<0$, using formulae (\ref{Newt-potential}) and (\ref{Phi-solution-everywhere}) we have
$1-\gamma(r) > 0$ for $r>R_\odot$.

The most stringent bound on $\gamma$ is given by the Cassini measurement \cite{Cassini}:
\begin{equation}\label{Cassini-bound}
\gamma - 1 = (2.1 \pm 2.3) \times 10^{-5},
\end{equation}
which implies
\begin{equation}\label{Cassini-upper-bound}
0 < 1 - \gamma < 2 \times 10^{-6}.
\end{equation}
Using then (\ref{Phi-solution-everywhere}) and (\ref{gamma-expression}), and neglecting $1-\gamma$ with respect to $1$ because of the Cassini bound
(\ref{Cassini-bound}), we get
\begin{equation}\label{1-gamma-formula}
(1-\gamma)\frac{U}{c^2} = -\Psi \left( 1 - \frac{1-\gamma}{2} \right) \approx -\Psi,
\end{equation}
from which, using  $U(r)=GM_\odot\slash r$ for the Newtonian potential for $r>R_\odot$ in the interplanetary space (neglecting the
contribution of the solar atmosphere), where $M_\odot$ is the Sun's mass, we obtain
\begin{equation}\label{1-gamma-result}
1 - \gamma \approx - \frac{c^2}{GM_\odot}\, r\Psi.
\end{equation}
Using now the solution (\ref{Psi-solution-solar-sys}) for $\Psi$, we have
\begin{equation}
\frac{d}{dr}\left( r\Psi(r) \right) < 0, \qquad\mbox{for } r_s<r<r_g,
\end{equation}
from which it follows that $1-\gamma(r)$ is an increasing function of $r$ in the interplanetary space.
For $r>r_g$, where the galactic potential $U_g$ dominates over the Newtonian potential of the Sun (both approaching zero at infinity),
the potential profile of the galaxy varies slowly enough in such a way that, for or purposes, it can be considered  
constant over the solar neighbourhood. Then, using (\ref{Psi-outer-solution}) and (\ref{1-gamma-formula}) with $U=U_g$,
it follows that $1-\gamma(r)$ is a decreasing function of $r$ in this region. Such a qualitative behaviour of $1-\gamma(r)$ is in agreement
with the result found numerically in Ref. \cite{HS} for $f(R)$ gravity.

Inserting now the expression (\ref{Psi-solution-solar-sys}) for $\Psi$ in formula (\ref{1-gamma-result}) for $1-\gamma$, 
and using the upper bound (\ref{Cassini-upper-bound}) from the Cassini measurement,
we obtain the following inequality:
\begin{widetext}
 \begin{equation}
\frac{16\pi}{M_\odot} q_1 \rho_s^\prime r_s^2  + \frac{2}{3}\,\frac{M_{\rm eff}(r)}{M_\odot}
+ \frac{8\pi}{3}\,\frac{r}{M_\odot}\int_r^{r_g} \rho \left[ 1 - \left(\frac{\rho_g}{\rho}\right)^{-\alpha\slash(1-\alpha)} \right] r^\prime dr^\prime
< 2 \times 10^{-6},
\end{equation}
\end{widetext}
for $r$ varying in the interplanetary space. The integral is extended over the solar atmosphere and, using the corresponding values of mass density
reported in Appendix B, it turns out that the integral term is negligible in comparison with $10^{-6}$ so that, taking into account
the approximation (\ref{first-integral}) of the effective mass, the previous inequality is approximated by
 \begin{equation}\label{Cassini-total-constraint}
\frac{16\pi}{M_\odot} q_1\rho_s^\prime r_s^2 + \frac{2}{3}\,\frac{M_{\rm eff}(R_\odot)}{M_\odot}  < 2 \times 10^{-6}.
\end{equation}
Since the effective mass $M_{\rm eff}(R_\odot)$ depends on $r_s$, and the screening radius is
determined by the integral equation (\ref{rs-integr-equation}) which contains the NMC gravity parameters $\alpha,q_1,q_2$, then
inequality (\ref{Cassini-total-constraint}) is a constraint on NMC gravity parameters.

We observe that if $q_1=0$ inequality (\ref{Cassini-total-constraint}) holds true also for $f(R)$ gravity (see related results in Ref. \cite{HS}).

\subsection{Computation of the screening radius and of the effective mass}

The mass of the Sun's convection zone is about $0.02M_\odot$  \cite{Christ} and the mass of the solar atmosphere is about $10^{-10}M_\odot$.
If $\varepsilon$ in inequality (\ref{consist-cond-10-3}) is small enough ($\varepsilon\leq 10^{-2}$ is sufficient), then using inequalities
(\ref{zconv-first-ineq-eps}) and (\ref{Cassini-total-constraint}), it follows that
the screening radius that saturates the Cassini constraint, {\it i.e.} that makes the inequality (\ref{Cassini-total-constraint}) an equality,
lies in the convection zone.

Moreover, for such values of $\varepsilon$
inequality (\ref{Cassini-total-constraint}) is satisfied only if $R_\odot-r_s\ll R_\odot$, so that the contribution to the effective mass only comes
from a thin shell \cite{KW} of radii $(r_s,R_\odot)$ in the upper part of the convection zone and in the photosphere. 

In this section we estimate the screening radius in the convection zone, as a function of parameters $\alpha,q_1,q_2$,
by means of an approximate solution of the integral equation (\ref{rs-integr-equation}),
obtained by resorting to the solar density model reported in Appendix B.

In order to solve the integral equation (\ref{rs-integr-equation}) we use the approximation (\ref{second-integral}).
We denote $r_p$ the radius at the base of the photosphere, which coincides with the top of the convection zone.
If $r_s$ lies in the convection zone, since the integration is extended over the thin shell, then we approximate
the integral on the right hand side of (\ref{second-integral}), considered as a function of $r_s$,
by means of the second order Taylor expansion around $r_p$, obtaining
\begin{equation}\label{integr-Taylor-approx}
\frac{1}{6}\int_{r_s}^{R_\odot}\rho(r)rdr \approx I_{\rm ph} +  P(r_s),
\end{equation}
where $I_{\rm ph} = 4.19\times 10^{10}\,{\rm g\,cm^2}$
is the contribution of the photosphere computed by using the mass density profile reported in Section \ref{sec:photosphere} of Appendix B, and
\begin{equation}
P(r_s) = \frac{1}{12} \left[ \rho_p\left(r_p^2-r_s^2\right)
- \rho^\prime_p r_p\left( r_p-r_s \right)^2 \right],
\end{equation}
where  $\rho_p=\rho(r_p)$, $\rho_p^\prime=d\rho\slash dr(r_p)$.
The quantities $\rho_p$ and $\rho^\prime_p$ are computed by using the density profile of the convection zone reported in Section \ref{sec:convection}
of Appendix B, and the derivative is computed from the side of the convection zone.

The term $\rho_s+\rho_s^\prime r_s$ in the integral equation (\ref{rs-integr-equation}) is approximated in the thin shell by means of
the quadratic approximation of density inside the convection zone starting at $r_p$:
\begin{eqnarray}
\rho_s+\rho_s^\prime r_s &\approx& Q(r_s) = \rho_p - \rho_p^\prime (r_p-2r_s) \nonumber\\
&+& \frac{1}{2}\rho_p^{\prime\prime} (r_p-r_s)(r_p-3r_s).
\end{eqnarray}
Substituting such second order approximations in the integral equation (\ref{rs-integr-equation}), the screening
radius $r_s$ then solves the following quadratic equation:
\begin{equation}\label{rs-quadr-eq}
P(r_s) + q_1 Q(r_s) + I_{\rm ph} = \alpha q_2 \left(\frac{8\pi G}{c^2}\right)^{\alpha-1}\rho_g^\alpha.
\end{equation}
The solution $r_s=r_s(\alpha,q_1,q_2)$ has to be selected out of the two roots of the quadratic equation
and then substituted into inequality (\ref{Cassini-total-constraint}), particularly into
the expression of the effective mass.
The contribution of the thin shell to the effective mass is computed by means of an analogous 
second order Taylor approximation of the integral in (\ref{first-integral}):
\begin{eqnarray}\label{eff-mass-Taylor}
M_{\rm eff}(R_\odot) &\approx& 4\pi r_p\left( r_p-r_s \right)\left[ \rho_p r_s - \frac{1}{2}\rho^\prime_p r_p\left( r_p-r_s \right) \right] \nonumber\\
&+& M_{\rm ph},
\end{eqnarray}
where $M_{\rm ph} = 2.2\times 10^{23}\,{\rm g}$
is the mass of the photosphere computed by using the mass density profile reported in Section \ref{sec:photosphere} of Appendix B.

Eqs. (\ref{Cassini-total-constraint}), (\ref{rs-quadr-eq}) and (\ref{eff-mass-Taylor}) will be used in the next section to find the Cassini constraint on NMC gravity parameters
$\alpha,q_1,q_2$.

\subsection{Constraints on NMC gravity parameters}\label{sec:Cassini-exclus-plots}

The constraint from Cassini measurement determines an admissible region in the three-dimensional parameter space with coordinates $\alpha,q_1,q_2$.
We represent the admissible region by means of two-dimensional exclusion plots obtained using sections with planes $\alpha=constant$ and $q_1=constant$.

In order to avoid either too small or too large numbers we replace parameters $q_1,q_2$ with the following rescaled, dimensionless parameters:
\begin{equation}\label{q1q2-rescaled}
\widetilde q_1 = \frac{q_1}{R_\odot^2}, \qquad \widetilde q_2 = q_2 R_g^\alpha,
\end{equation}
with $R_g=8\pi G\rho_g\slash c^2$. With this substitution, the quadratic equation (\ref{rs-quadr-eq}) which determines the screening radius $r_s$ becomes
\begin{equation}
\frac{8\pi G}{c^2} \left[ P(r_s) + \widetilde q_1 R_\odot^2 Q(r_s) + I_{\rm ph} \right] = \alpha \widetilde q_2.
\end{equation}
The admissible root, which has to satisfy $r_s\leq r_p$, expressed in the form $r_s=r_s(\alpha,\widetilde q_1,\widetilde q_2)$,
is substituted in the approximation (\ref{eff-mass-Taylor}) of the effective mass, then the admissible values of parameters $\alpha,\widetilde q_1,\widetilde q_2$ 
satisfy inequality (\ref{Cassini-total-constraint}).
The admissible region in parameter space is restricted by means of the condition $\lambda_g\gg r_g$ introduced in Section \ref{Sec:outskirts} which,
expressed in terms of parameters $\alpha,\widetilde q_2$ and using Eq. (\ref{lambda-expression}) with $\rho=\rho_g$, becomes
\begin{equation}
\label{q2lowerbound}
\left[ \frac{3}{4\pi}\alpha(1-\alpha)\frac{c^2}{G\rho_g}\widetilde q_2 \right]^{1\slash 2} > 10^2 r_g,
\end{equation}
where we have required $\lambda_g>10^2 r_g$. The admissible region is further restricted by the consistency condition, inequality (\ref{consist-cond-10-3}):
\begin{equation}\label{consistency-condition-rewritten}
|q_1| < \varepsilon \, \frac{\rho(r)}{|\nabla^2\rho(r)|} \quad \mbox{for any } r<r_s.
\end{equation}
Using the expression of $\nabla^2\rho$ in Section \ref{sec:convection} of Appendix B, the ratio $\rho/|\nabla^2\rho|$ is a decreasing function of $r$ in the convection zone,
so that, for given $\varepsilon$, the upper bound on $|q_1|$ is smaller for $r=r_s$. Moreover, for given values of $q_1$ and $\varepsilon$ the consistency condition yields an upper
bound $r_s< r_s^\ast=r_s^\ast(q_1,\varepsilon)$ on the screening radius, so that, using the integral equation (\ref{rs-integr-equation}), the approximation (\ref{second-integral}), 
and inequality (\ref{zconv-second-ineq-eps}), we get the following lower bound on $\alpha \widetilde q_2$ for $\varepsilon\leq 1/20$:
\begin{equation}\label{alpha-q2-low-bound}
\alpha\widetilde q_2 > \frac{4\pi G}{3c^2}\left[ (1-20\varepsilon)\int_{r_s^\ast}^{R_\odot}\rho(r)rdr \right].
\end{equation}

\begin{widetext}
	
\begin{figure}[H]
\centering
\begin{minipage}{.49\columnwidth}
	\centering
	\includegraphics[width=1.01\textwidth]{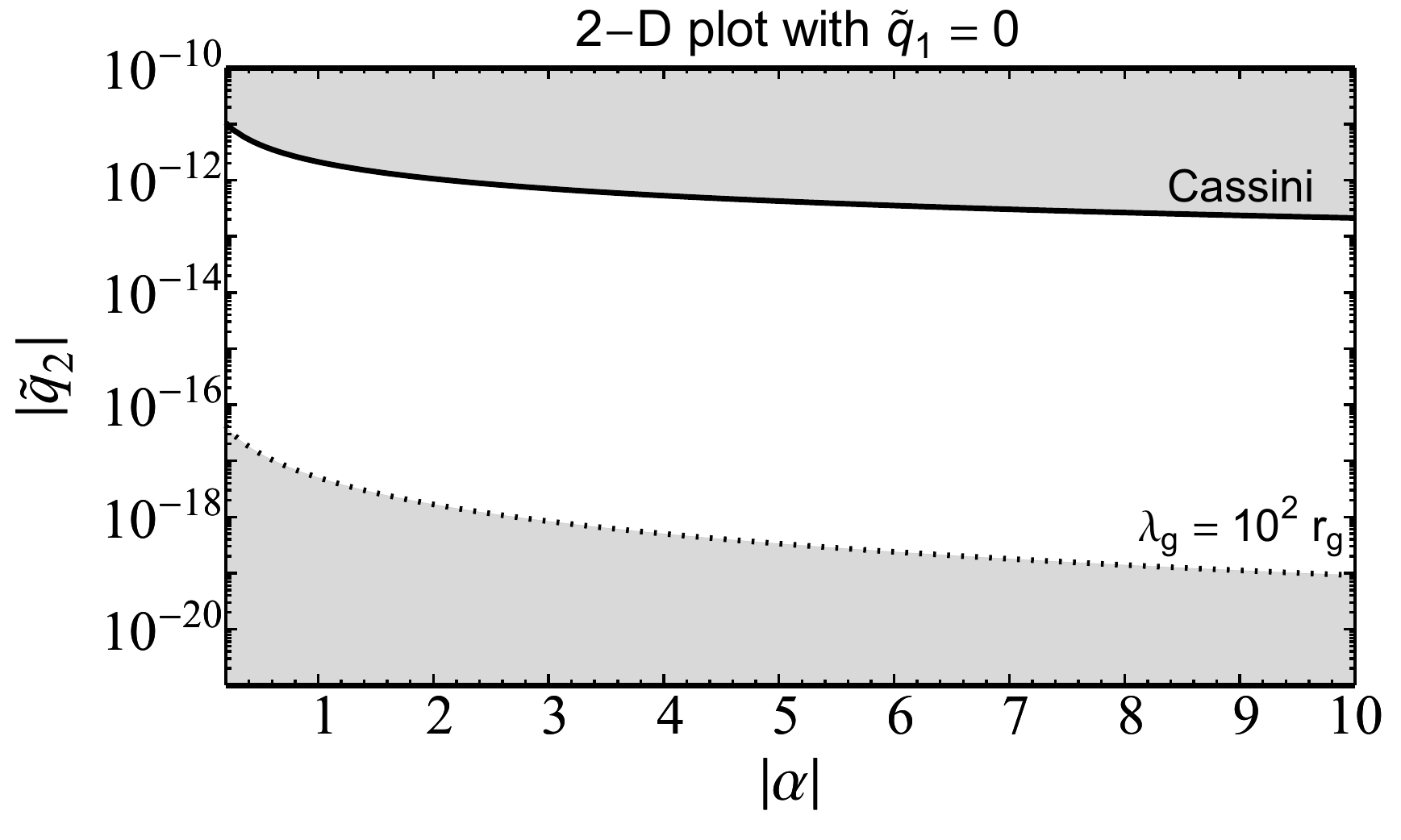}
	\caption{Cassini constraint on the parameter quarter plane $|\alpha|,|\widetilde q_2|$ for $\widetilde q_1=0$. The solid line yields the upper bound on $|\widetilde q_2|$
	from Cassini measurement, the dotted line yields the lower bound on $|\widetilde q_2|$ from inequality $\lambda_g > 10^2 r_g$.}
	\label{fig:alpha-q2-q1=0}
\end{minipage}\hfill
\begin{minipage}{.49\columnwidth}
	\centering
	\includegraphics[width=1.01\textwidth]{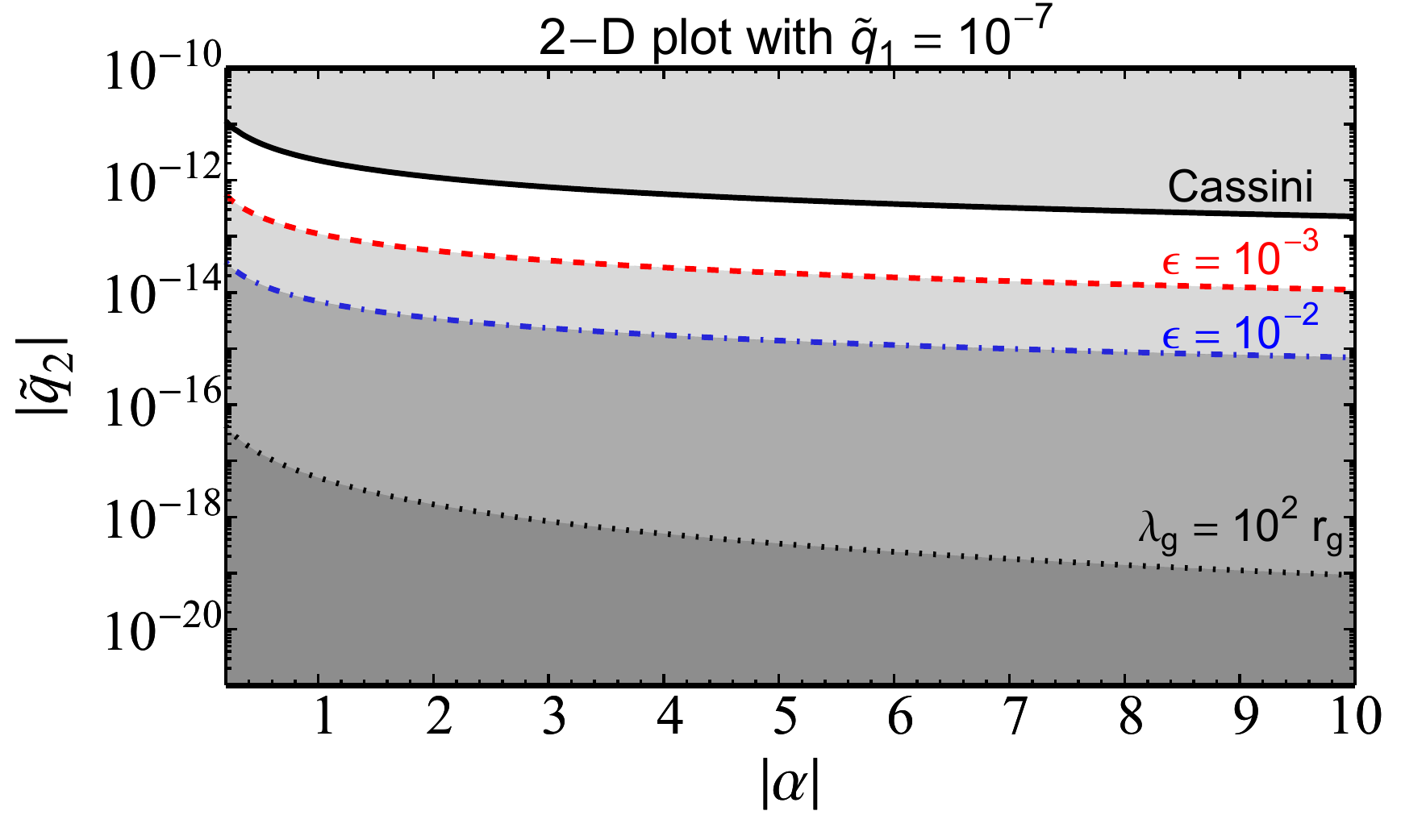}
	\caption{Cassini constraint on the parameter quarter plane $|\alpha|,|\widetilde q_2|$ for $\widetilde q_1=10^{-7}$. The dotted colored lines yield the lower bound
	on $|\widetilde q_2|$ from inequality (\ref{alpha-q2-low-bound}). If $\varepsilon=10^{-2}$ the zone plotted in light grey between the dotted
	colored lines is admissible.}
	\label{fig:alpha-q2-q1=10-7}
\end{minipage}
\end{figure}
\end{widetext}

\begin{widetext}
	
\begin{figure}[H]
\centering
\begin{minipage}{.49\columnwidth}
	\centering
	\includegraphics[width=1.01\textwidth]{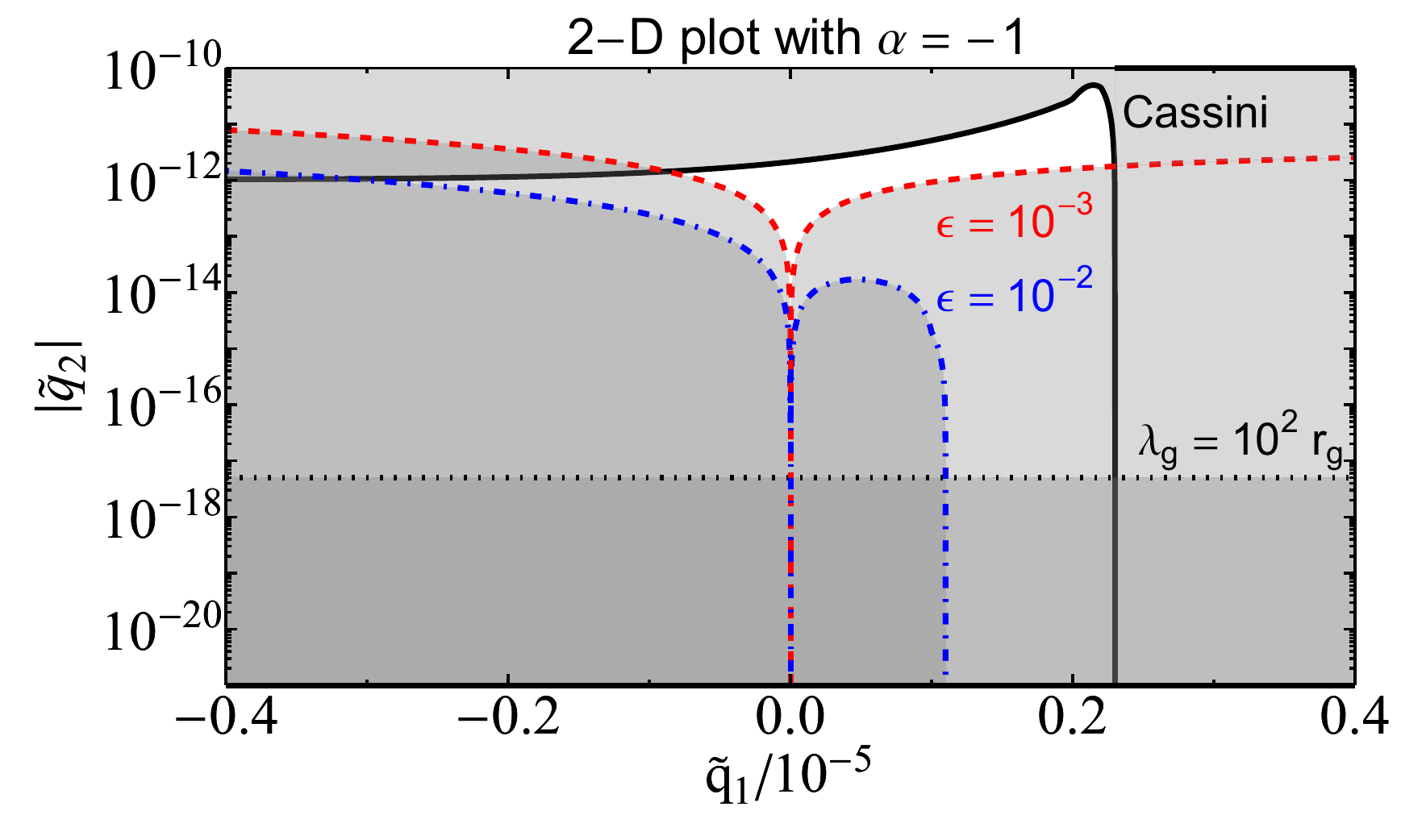}
	\caption{Cassini constraint on the parameter half plane $\widetilde q_1,|\widetilde q_2|$ for $\alpha=-1$. The solid line is the Cassini bound,
	the dotted black line is the bound from inequality $\lambda_g > 10^2 r_g$, the dotted colored lines are the bounds
	from inequality (\ref{alpha-q2-low-bound}). If $\varepsilon=10^{-2}$ the zone plotted in light grey between the dotted colored lines and the Cassini bound is admissible.}
	\label{fig:q1-q2-alpha=-1}
\end{minipage}\hfill
\begin{minipage}{.49\columnwidth}
	\centering
	\includegraphics[width=1.01\textwidth]{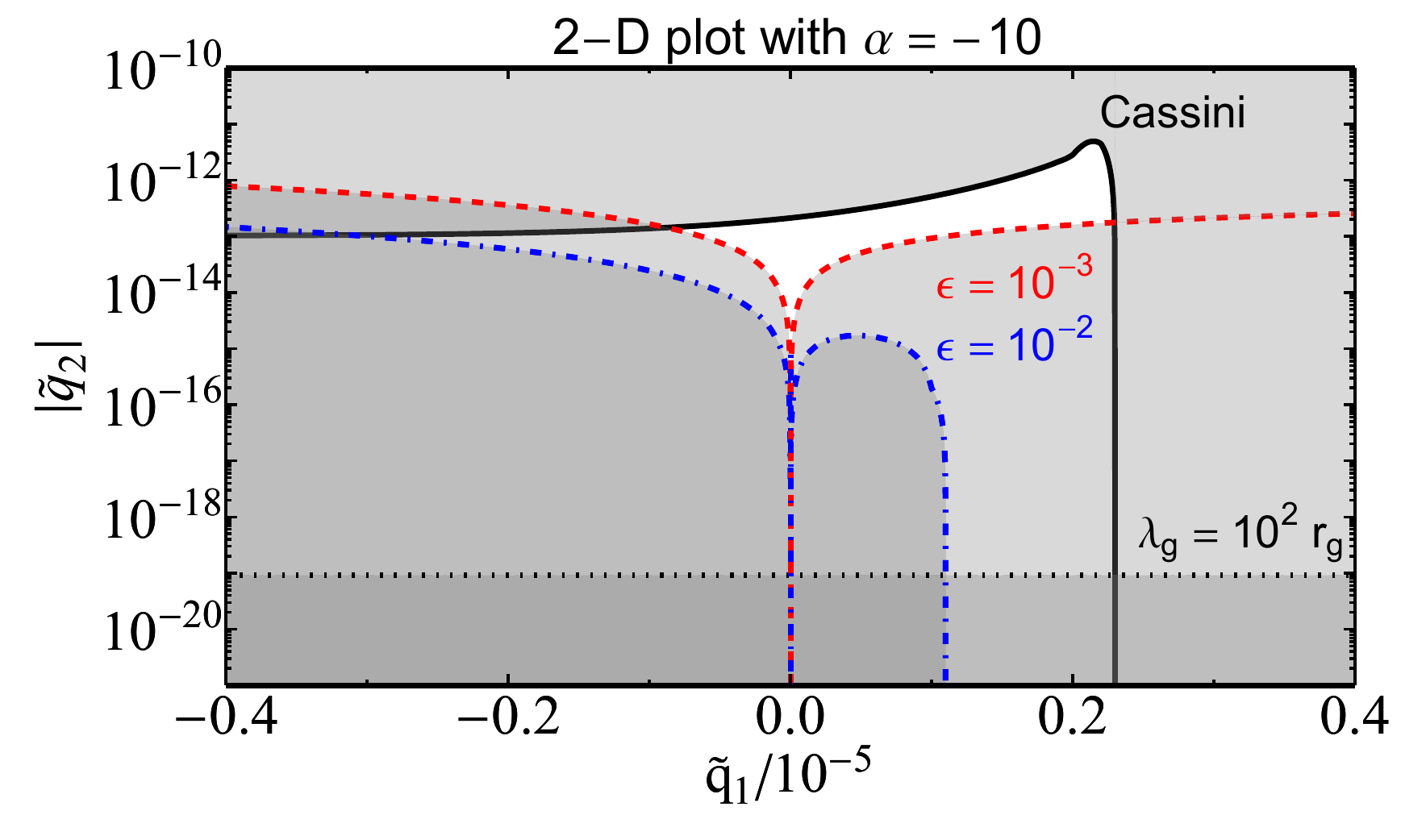}
	\caption{Cassini constraint on the parameter half plane $\widetilde q_1,|\widetilde q_2|$ for $\alpha=-10$. The solid line is the Cassini bound,
	the dotted black line is the bound from inequality $\lambda_g > 10^2 r_g$, the dotted colored lines are the bounds
	from inequality (\ref{alpha-q2-low-bound}). If $\varepsilon=10^{-2}$ the zone plotted in light grey between the dotted colored lines and the Cassini bound is admissible.}
	\label{fig:q1-q2-alpha=-10}
\end{minipage}
\end{figure}
\end{widetext}

\begin{widetext}
\begin{figure}[h!]
\centering
\includegraphics[scale=1]{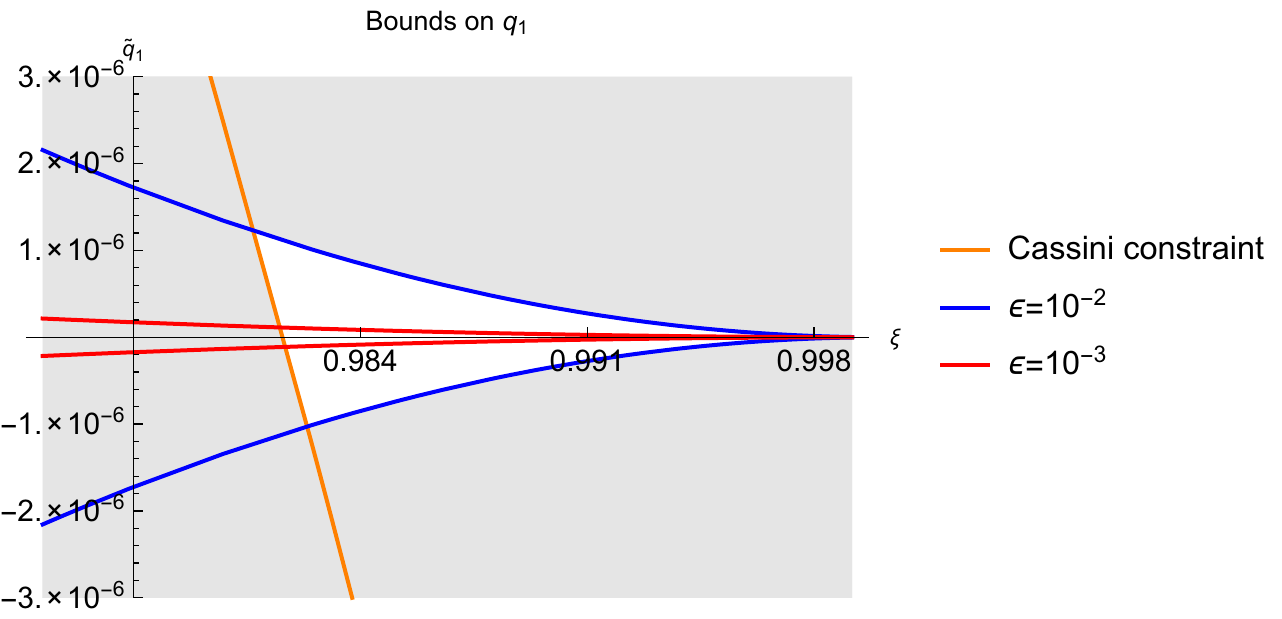}
\caption{Allowed values of $\widetilde q_1$ as function of the dimensionless screening radius $\xi=r_s\slash R_\odot$, for different values of the parameter $\varepsilon$ from the consistency condition, Eq. (\ref{consist-cond-10-3}). The excluded regions are colored in grey.}
\label{fig:q1bounds}
\end{figure}
\end{widetext}

\begin{widetext}
	
\begin{figure}[H]
\centering
\begin{minipage}{.49\columnwidth}
	\centering
	\includegraphics[width=1.01\textwidth]{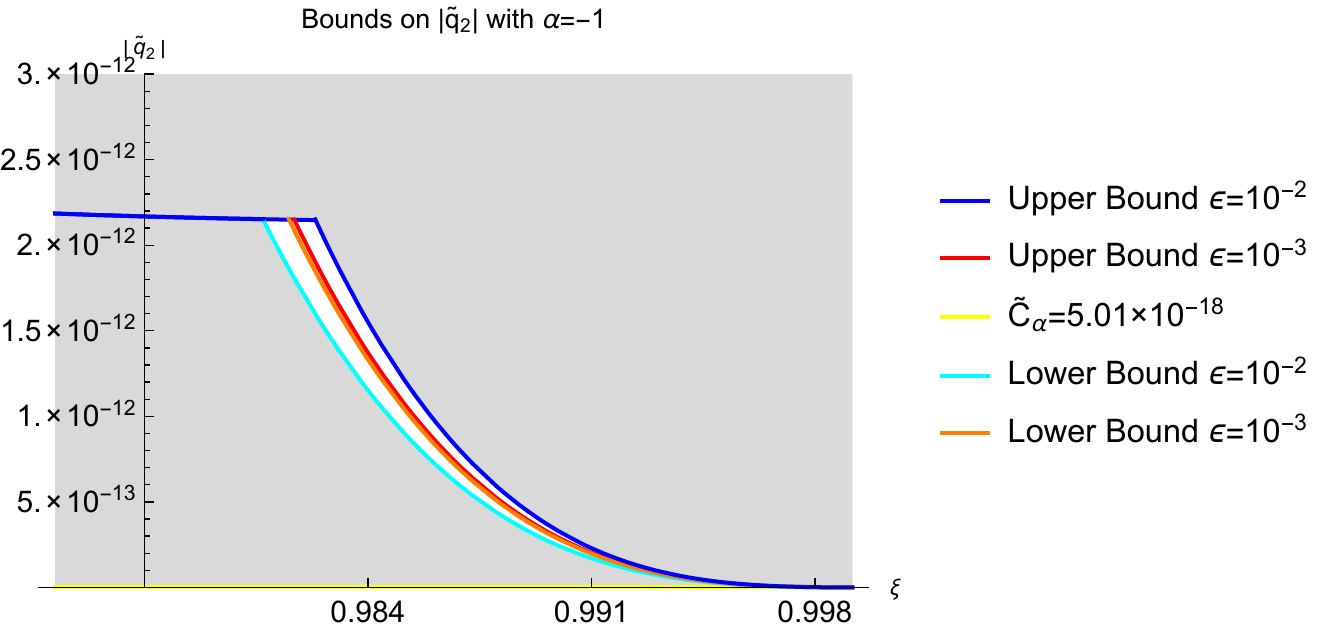}
	\caption{Allowed values of $|\widetilde q_2|$ for $\alpha=-1$ as function of the dimensionless screening radius $\xi=r_s\slash R_\odot$ for different values of the parameter $\varepsilon$ from the consistency condition, Eq. (\ref{consist-cond-10-3}). The excluded regions are colored in grey.}
	\label{fig:rs-q2-alpha=-1}
\end{minipage}\hfill
\begin{minipage}{.49\columnwidth}
	\centering
	\includegraphics[width=1.01\textwidth]{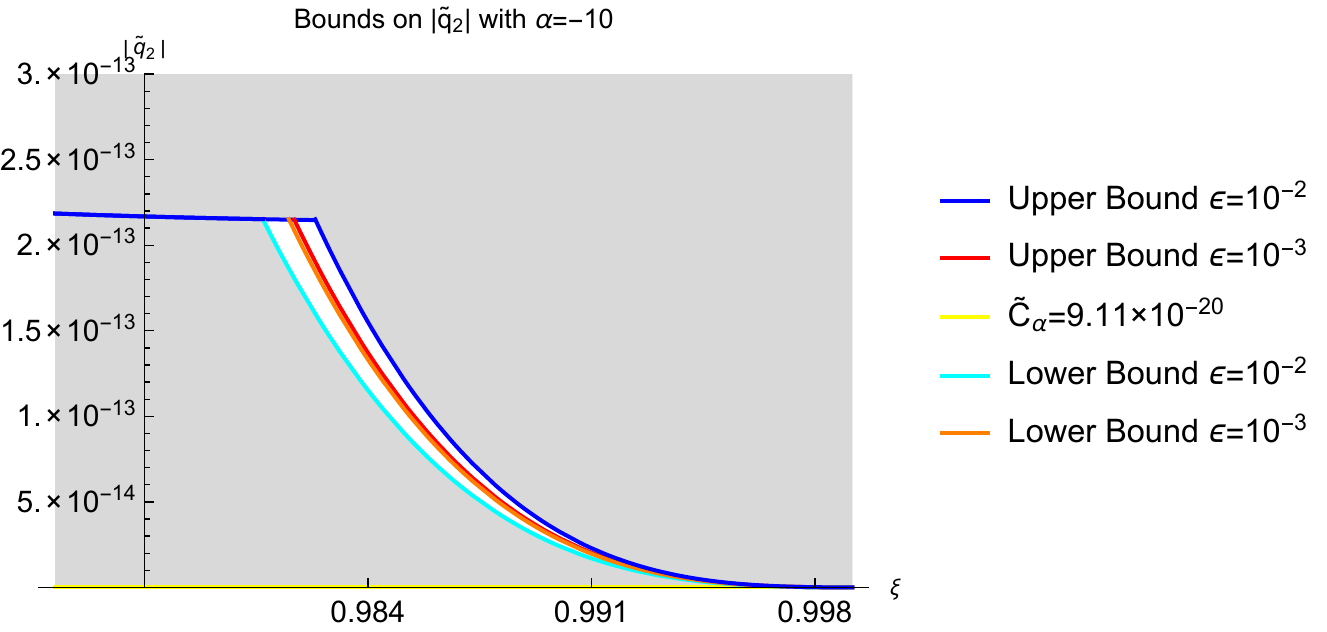}
	\caption{Allowed values of $|\widetilde q_2|$ for $\alpha=-10$ as function of the dimensionless screening radius $\xi=r_s\slash R_\odot$ for different values of the parameter $\varepsilon$ from the consistency condition, Eq. (\ref{consist-cond-10-3}). The excluded regions are colored in grey.}
	\label{fig:rs-q2-alpha=-10}
\end{minipage}
\end{figure}
\end{widetext}

Our results are graphically reported in Figures \ref{fig:alpha-q2-q1=0}-\ref{fig:rs-q2-alpha=-10}: admissible regions for parameters are plotted in white,
while the excluded regions are plotted in grey. 
Fig. \ref{fig:alpha-q2-q1=0} shows the section of the admissible region in three-dimensional parameter space
with the plane $\widetilde q_1=0$; since both $\alpha$ and $\widetilde q_2$ are negative the admissible region is plotted in the quarter plane with coordinates
$(|\alpha|,|\widetilde q_2|)$.
Fig. \ref{fig:alpha-q2-q1=10-7} shows the sections with the plane $\widetilde q_1=10^{-7}$ corresponding to values $\varepsilon=10^{-2}$
and $\varepsilon=10^{-3}$.  Using inequality (\ref{consistency-condition-rewritten}) the value $\widetilde q_1=10^{-7}$ corresponds to the upper bound
$r_s^\ast\approx 0.99R_\odot$ for $\varepsilon=10^{-2}$. Fig. \ref{fig:alpha-q2-q1=10-7} shows that the lower bound (\ref{alpha-q2-low-bound}) is greater than the one resulting from inequality
$\lambda_g>10^2 r_g$.

Figures \ref{fig:q1-q2-alpha=-1} and \ref{fig:q1-q2-alpha=-10} show the sections of the admissible region in three-dimensional parameter space
with the planes $\alpha=-1$ and $\alpha=-10$, respectively. The admissible region is plotted in the half plane with coordinates $(\widetilde q_1,|\widetilde q_2|)$.
In the case $\widetilde q_1=0$, the value $\alpha=-1$ has been used in Ref. \cite{drkmattgal} to model the rotation curves of galaxies, and the value $\alpha=-10$
has been used in Ref. \cite{curraccel} to model the current accelerated expansion of the Universe. The asymmetry of the admissible region with respect to the
axis $\widetilde q_1=0$ is due to the impact of the sign of $q_1$ in the solution of the integral equation (\ref{rs-integr-equation}).

Figure \ref{fig:q1bounds} shows the allowed region for $\widetilde q_1$ as function of the dimensionless screening radius $\xi=r_s\slash R_\odot$.
That is computed by intersecting the region allowed by the inequality of the Cassini constraint,  Eq. (\ref{Cassini-total-constraint}), with
the region allowed by the consistency condition, Eq. (\ref{consist-cond-10-3}), for the values $\varepsilon=10^{-2}$ and $\varepsilon=10^{-3}$.
Since $\rho_s^\prime<0$ the Cassini constraint yields a lower bound on $\widetilde q_1$ while the consistency condition yields both an upper (positive) bound on $\widetilde q_1$ and a
lower (negative) bound. The intersection point between the curve of the Cassini constraint and the curve of the upper bound from the consistency condition, defines
a lower bound on the screening radius. Such a lower bound $\xi_{\rm low}$ is the value of the screening radius that saturates the Cassini constraint: values of
$r_s$ below $\xi_{\rm low}R_\odot$ are excluded. Figure \ref{fig:q1bounds} shows that the lower bound on $\xi$ weakly depends on
$\varepsilon$ and its value is $\xi_{\rm low}\approx 0.98$ which corresponds to a maximum radial thickness of the thin shell of about $0.02R_\odot$.

Figures \ref{fig:rs-q2-alpha=-1} and \ref{fig:rs-q2-alpha=-10} show the allowed region for $|\widetilde q_2|$ as function of $\xi$ for a fixed value of $\alpha$.
That is computed by solving the integral equation (\ref{rs-integr-equation}) with respect to $q_2$, then allowing $q_1$ to vary in its admissible region.
This has been done for $\alpha=-1$ in Fig. (\ref{fig:rs-q2-alpha=-1}), and $\alpha=-10$ in Fig. \ref{fig:rs-q2-alpha=-10}; in both cases the values $\varepsilon=10^{-2}$ and $\varepsilon=10^{-3}$
have been used. The constant $\widetilde C_{\alpha}$ corresponds to the lower bound on $\widetilde q_2$ from inequality (\ref{q2lowerbound}), and both figures show that
its value is much smaller than the lower bound computed by means of the integral equation.
Eventually, we observe that $|\widetilde q_2|$ decreases monotonically as the screening radius increases.

\subsection{Verification of the consistency condition}

Once the constraint from Cassini measurement has been satisfied, we have to check that the consistency condition (\ref{consist-condition}) is verified in the solar interior for $r\leq r_s$.
We verify such a condition in the convection zone: using the model of density profile $\rho(r)$ in this zone, reported in Section \ref{sec:convection} of Appendix B, we have
\begin{equation}
\nabla^2\rho = \frac{n_c-1}{n_c\rho}\left(\frac{d\rho}{dr}\right)^2, \qquad n_c=2.33,
\end{equation}
from which, using Eq. (\ref{lambda-rescaled-lambdag}) for $\lambda^2$ and taking into account that $\nabla^2\rho(r)>0$ in the convection zone, the consistency condition becomes
\begin{equation}\label{cons-cond-substituted}
\frac{\nabla^2\rho}{\rho}\left\vert \lambda_g^2\left(\frac{\rho}{\rho_g}\right)^{\alpha-1}\left[ \frac{\alpha}{1-\alpha}-\frac{n_c}{n_c-1}(\alpha+1) \right] + 6q_1 \right\vert \ll 1.
\end{equation}
An expression for $\lambda_g$ can be found from the general expression (\ref{lambda-expression}) of $\lambda$ and the integral equation (\ref{rs-integr-equation}).
Let us choose $|q_1|<\varepsilon 10^{-7}R_\odot^2$, with $\varepsilon<10^{-2}$ (see Section \ref{sec:eta-sol-interior}).
Then, using inequality (\ref{zconv-second-ineq-eps}), for an estimate of the order of magnitude of $\lambda_g$ we can neglect the term involving $q_1$ in the integral equation, obtaining:
\begin{equation}\label{lambg-integral}
\lambda_g^2\approx \frac{1-\alpha}{\rho_g}\int_{r_s}^{R_\odot}\rho(r)rdr,
\end{equation}
where we have neglected the contribution to the integral from density in the solar atmosphere outside the photosphere.
Now, it turns out that both
\begin{equation}
\frac{\nabla^2\rho}{\rho} \qquad\mbox{and}\qquad \left(\frac{\rho}{\rho_g}\right)^{\alpha-1}
\end{equation}
are increasing functions of $r$ in the convection zone, and the maximum of $\nabla^2\rho\slash\rho$ is of order of $10^7\slash R_\odot^2$.
Hence, if the consistency condition is satisfied for $r=r_s$, then it is satisfied also for any $r< r_s$ and
the verification of such a condition depends mainly on the factor $(\rho\slash\rho_g)^{\alpha-1}$.

Let us now choose for instance the value $r_s\approx 0.98R_\odot$ for the screening radius, which
saturates the Cassini constraint (see the results in the previous section).
Using the density profile in Section \ref{sec:convection} of Appendix B, we have
\begin{equation}
\frac{\nabla^2\rho(r)}{\rho(r)} < \frac{10^7}{R_\odot^2}, \quad \left(\frac{\rho(r)}{\rho_g}\right)^{\alpha-1} < 10^{-20(1-\alpha)}, \quad\mbox{for }r<r_s,
\end{equation}
and using (\ref{lambg-integral}) we have $\lambda_g^2\approx(1-\alpha)10^{18}R_\odot^2$. 
Inserting these numbers into inequality (\ref{cons-cond-substituted}) it turns out that the consistency condition is verified for the values $\alpha\leq -1$ which satisfy the Cassini constraint.

Analogous results can be found for different values of $r_s$ (satisfying the Cassini constraint) and for $r$ varying in the radiative interior.

\section{Effect of the extra force in the solar atmosphere}\label{sec:extra-f-atmosph}

We compute the effect of fifth force and extra force on temperature and density in the solar atmosphere when the screening radius lies
in the convection zone. We consider the chromosphere-corona transition region where a steep density gradient takes place \cite{Mariska}
(see Section \ref{sec:chromo-coro-transr} of Appendix B), and the extra force is expected to have largest intensity. 

Models of the transition region have been derived from the distribution of the emission measure
computed from the intensities of spectral lines in the ultraviolet region of the solar spectrum \cite{Jordan76}.
In the approximation of a plane parallel geometry the average emission measure distribution $EM$, for the quiet Sun, 
is parametrized as a function of temperature $T$ by \cite{Jordan}
\begin{equation}\label{emission-measure}
EM = \int_{\Delta T} N_e^2 dh = a T^b,
\end{equation}
where $N_e$ is the electron number density, $h=h(T)$ is height above the solar limb, 
$\Delta T$ denotes the temperature range $\Delta T = (0.891T,1.122T)$ used in \cite{RayDoyle},
and $a,b$ are parameters which take different values in the two zones below and above the temperature $T_0=1.5\times 10^5\,{\rm K}$, respectively.

In the sequel we use the following approximations: we assume hydrostatic equilibrium, complete ionization of hydrogen (which is realized for $T\geq 2\times 10^4$ K),
single ionization of helium in the zone below $T_0$ (lower transition region), and double ionization of helium in the zone above $T_0$ (upper transition region).
Moreover, we neglect elements other than hydrogen and helium both in the equation of hydrostatic equilibrium
and in ionization equilibrium. Assuming the perfect gas law, the electron density is then given by $N_e=p\slash \xi k_B T$, where $p$ is pressure, $k_B$ is the Boltzmann constant,
and $\xi$ is related to the degree of ionization of helium: $\xi= 2$ in the lower transition region, and $\xi= 1.91$ in the upper transition region.
Equality (\ref{emission-measure}) is then written in the form
\begin{equation}\label{EM-integral}
\frac{1}{\xi^2 k_B^2}\int_{\Delta T} \frac{p^2}{(T^\prime)^2}\frac{dh}{dT^\prime}dT^\prime = a T^b.
\end{equation}
\smallskip

\noindent
Following the method in \cite{Jordan76,Jordan}, both pressure $p$ and
the quantity $(1/T^2)dh\slash dT$ are approximated with a constant over the temperature interval $\Delta T$, so that Eq. (\ref{EM-integral}) yields
for a monotone temperature profile $T(h)$:
\begin{equation}\label{EM-explicited}
\frac{p^2}{T} \approx \frac{\xi^2}{0.231}k_B^2 \frac{a}{b+1} \frac{d}{dh}\left( T^{b+1} \right),
\end{equation}
where, both in this formula and in the following computations, the two zones in the transition region below and above the temperature $T_0$, have to be considered
separately (see Section \ref{sec:chromo-coro-transr} of Appendix B).

Multiplying now by pressure $p$ both sides of Eq. (\ref{NMC-hydro}) in the case of hydrostatic equilibrium ({\it i.e.} $dv/dt=0$), 
approximating $dU/dr\approx -GM_\odot/R_\odot^2$, replacing $d/dr$ with $d/dh$, using $\rho=\mu m_p p/k_BT$ with $m_p$ the proton mass and $\mu$ the mean molecular weight,
and using expressions (\ref{Fifth-force-outside-rs}) with $r^2\approx R_\odot^2$ and (\ref{extra-force-outside-rs}) of fifth force and extra force outside the screening radius, 
respectively, we obtain
\begin{eqnarray}\label{p-by-hydro-eq}
\frac{d}{dh}\left(\frac{p^2}{2}\right) &=& -\frac{\mu m_p G}{k_B R_\odot^2}\left(M_\odot + \frac{1}{3}M_{\rm eff}(R_\odot) + 8\pi q_1\rho^\prime_s r_s^2 \right)\frac{p^2}{T} \nonumber\\
&+& \mathcal{F}(\alpha,q_1,q_2) \, p \rho^{\frac{\alpha}{1-\alpha}} \,\frac{d\rho}{dh}.
\end{eqnarray}
We use $\mu=0.65$ in the lower transition region and $\mu=0.62$ in the upper transition region.
We denote $p_0$ the pressure at the temperature $T_0$, then setting $p=p_0+\delta p$, it is known that $\delta p$ is a small variation in the transition zone \cite{Doschek}, 
and we approximate
\begin{equation}
\mathcal{F}(\alpha,q_1,q_2) \, p \approx \mathcal{F}(\alpha,q_1,q_2) \, p_0,
\end{equation}
where we have negected the product $\mathcal{F}\delta p$. Then, combining Eqs. (\ref{EM-explicited}) and (\ref{p-by-hydro-eq}), 
and expressing again $\rho$ in terms of pressure and temperature, we find that the following expression has a vanishing derivative with respect to $h$:
\begin{widetext}
\begin{equation}\label{EM-vanish-deriv}
\frac{p^2}{2} + \frac{\xi^2}{0.231}\frac{\mu m_p k_B a}{b+1} \frac{G}{R_\odot^2}\left(M_\odot + \frac{1}{3}M_{\rm eff}(R_\odot) + 8\pi q_1\rho^\prime_s r_s^2 \right)T^{b+1}
- (1-\alpha)\mathcal{F}(\alpha,q_1,q_2)\left(\frac{\mu m_p}{k_B}\right)^{\frac{1}{1-\alpha}}p_0^{\frac{2-\alpha}{1-\alpha}}T^{\frac{1}{\alpha-1}} = constant,
\end{equation}
\end{widetext}
where we have again approximated $\mathcal{F}p^{1/1-\alpha}\approx\mathcal{F}p_0^{1/1-\alpha}$.
The value of the constant in Eq. (\ref{EM-vanish-deriv}) is given by the above expression evaluated at $T_0,p_0$,
so that pressure $p$ and electron number density $N_e$ can be computed, obtaining
\begin{equation}\label{electron-density-formula}
N_e(\alpha,q_1,q_2) = \frac{1}{\xi T}\left( A_1 + A_2 T^{b+1} + A_3 T^{\frac{1}{\alpha-1}} \right)^{1\slash 2},
\end{equation}
where we stress the dependence of $N_e$ on NMC parameters, and
\begin{eqnarray}
A_1 &=& \left(\frac{p_0}{k_B}\right)^2 - A_2 T_0^{b+1} - A_3 T_0^{\frac{1}{\alpha-1}}, \nonumber\\
A_2 &=& -\frac{8.658\xi^2}{b+1} a \frac{\mu m_pG}{k_BR_\odot^2}\left(M_\odot + \frac{1}{3}M_{\rm eff}(R_\odot) + 8\pi q_1\rho^\prime_s r_s^2 \right), \nonumber\\
A_3 &=& (1-\alpha)\frac{2}{k_B^2}\left(\frac{\mu m_p}{k_B}\right)^{\frac{1}{1-\alpha}}p_0^{\frac{2-\alpha}{1-\alpha}}\mathcal{F}(\alpha,q_1,q_2).
\end{eqnarray}
For pressure $p_0$ we use $p_0\slash k_B=2\times 10^{15}\,{\rm K\,cm^{-3}}$ derived from \cite{RayDoyle}, while the values of the various constants
are reported in Appendix B.

Accurate determination of electron density in the transition region is achieved by the observation of the intensity ratio of spectral lines emitted by the same ion 
at the temperature of formation of such an ion \cite{Doschek}. Particularly, the quiet Sun intensity ratios of lines emitted by the ${\rm Si}^{+2}$ ion
have been used to determine the electron density $N_e$ in the lower transition region \cite{Doschek}, where $T<T_0$, density exhibits the steepest gradient
(see Section \ref{sec:chromo-coro-transr} of Appendix B), and the extra force has largest intensity.

Uncertainties in the derived electron densities \cite{Doschek} are due to uncertainties in line ratios (typically less than $20\%$ for quiet Sun), spectral line blending
($25\%$-$30\%$ as maximum possible effect), time variations of the physical conditions in the solar atmosphere ($15\%$ change in line intensity for quiet Sun
over a time interval of 1.5 hr), and uncertainties in the atomic physics data (systematic): an overall uncertainty of a factor of 2 in density is then considered in \cite{Doschek}.

Since the measurement of electron density from spectral line ratios does not depend on gravity, while the expression (\ref{electron-density-formula}) of $N_e$
is affected by gravity, then equating such an expression with density $N_{e,{\rm obs}}$, observed from line ratios, yields a constraint on $\alpha,q_1,q_2$.

The electron density at the temperature of formation of the ${\rm Si}^{+2}$ ion in ionization equilibrium,
determined from intensity ratios of lines within the Si III multiplet near 1300 \AA, is found to be for quiet Sun \cite{Doschek}
\begin{equation}
N_{e,{\rm obs}} = 4.3 \times 10^{10} \, {\rm cm}^{-3}.
\end{equation}
Since $\mathcal{F}(\alpha,q_1,q_2)<0$, taking into account the overall density uncertainty considered above, we get the constraint
\begin{equation}\label{electron-density-constraint}
N_e(\alpha,q_1,q_2) \geq 2.15 \times 10^{10} \, {\rm cm}^{-3},
\end{equation}
where formula (\ref{electron-density-formula}) has to be evaluated using the temperature $T=3.2\times 10^4$ K of formation of the ${\rm Si}^{+2}$ ion,
and using the values of $a,b$ reported in Section \ref{sec:chromo-coro-transr} of Appendix B for the lower transition region.

We observe that, using the constraint (\ref{Cassini-total-constraint}) from Cassini measurement, the coefficient $A_2$ can be approximated by the Newtonian value
\begin{equation}\label{A2-Cassini-approxim}
A_2 \approx -\frac{8.658\xi^2}{b+1} a \frac{\mu m_p}{k_B}\,\frac{GM_\odot}{R_\odot^2},
\end{equation}
and, using the expression (\ref{Fifth-force-outside-rs}) of the fifth force outside the screening radius, it follows
\begin{equation}
\left\vert F_{\rm f}(r)\right\vert < 10^{-6}\rho(r)\frac{GM_\odot}{R_\odot^2},
\end{equation}
so that the fifth force is negligible in comparison with the Newtonian force.

\subsection{Constraints on NMC gravity parameters}\label{sec:extra-force-exclus-plots}

The constraint from spectroscopic measurements determines an admissible region in the three-dimensional parameter space with coordinates $\alpha,q_1,q_2$.
We represent the admissible region by means of two-dimensional exclusion plots obtained using sections with planes $\alpha=constant$.
We use the variables $\xi,\widetilde q_1,\widetilde q_2$ introduced in Section \ref{sec:Cassini-exclus-plots} about the Cassini constraint.

We find convenient to express the electron density $N_e$ as a function of $\alpha$, $q_1$ and the screening radius $r_s$, moreover, we use
approximation (\ref{A2-Cassini-approxim}). The screening radius appears in coefficient $A_3$, particularly in the term $\eta(R_\odot)$ inside $\mathcal{F}$,
both explicitly and implicitly through the effective mass $M_{\rm eff}(R_\odot)$.
In order to obtain the function $N_e=N_e(\alpha,q_1,r_s)$ we eliminate $q_2$ from $\mathcal{F}(\alpha,q_1,q_2)$ using the integral equation (\ref{rs-integr-equation}):
\begin{equation}
\alpha q_2 \approx \frac{1}{\rho_g^\alpha}\left(\frac{8\pi G}{c^2}\right)^{1-\alpha} \left[ q_1\left(\rho_s+\rho_s^\prime r_s\right) + \frac{1}{6}\int_{r_s}^{R_\odot}\rho(r)rdr\right],
\end{equation}
where we have again neglected the contribution to the integral from density in the solar atmosphere outside the photosphere. Then we substitute the above expression
of $\alpha q_2$ in formula (\ref{cal-F-eforce-approxim}) of $\mathcal{F}$ and in formula (\ref{eta-convz-simple}) of $\eta$,
which has to be evaluated in $r=R_\odot$ to compute $\mathcal{F}$. After these substitutions  the constraint (\ref{electron-density-constraint}) becomes
\begin{equation}
N_e(\alpha,q_1,r_s) \geq 2.15 \times 10^{10} \, {\rm cm}^{-3}.
\end{equation}
For given $\alpha$ such an inequality determines an admissible region in the half plane with coordinates $(q_1,r_s)$.
Such a region is further restricted by the consistency condition, inequality (\ref{consist-cond-10-3}), that has to be satisfied for any $r<r_s$.
Hence, using coordinates $(\widetilde q_1,\xi)$ in the half plane, the admissible region is restricted by means of the intersection with a region, symmetric with
respect to the axis $\widetilde q_1=0$, which shrinks as $\xi$ increases, being $\rho\slash\nabla^2\rho$ a decreasing function of $\xi$.

\begin{widetext}
	
\begin{figure}[H]
\centering
\begin{minipage}{.49\columnwidth}
	\centering
	\includegraphics[width=1.01\textwidth]{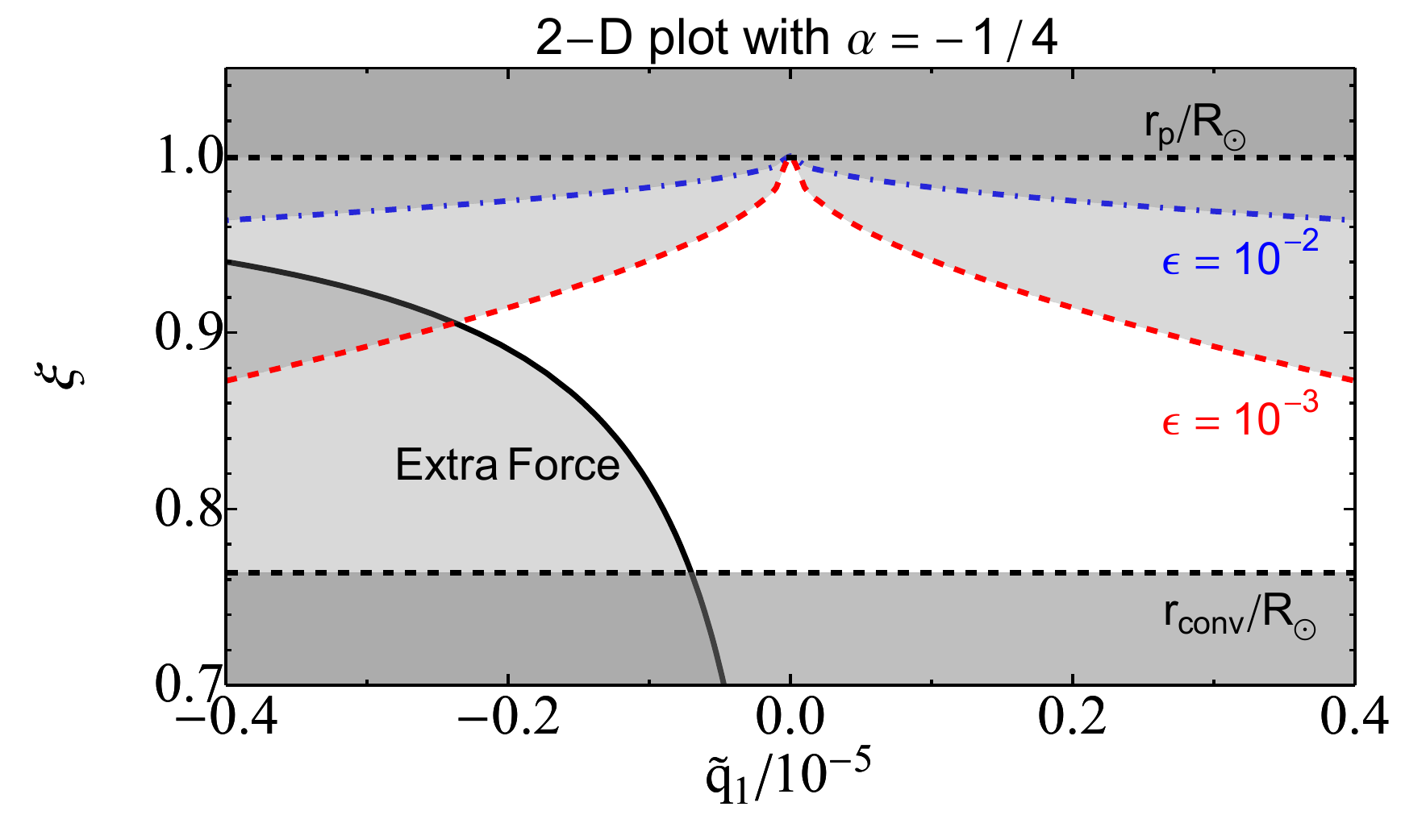}
	\caption{Constraint from extra force on the half plane $\widetilde q_1,\xi$ for $\alpha=-1/4$. The solid line yields the lower bound on the screening radius
	from spectroscopic measurements. The dotted colored lines yield the bound from the consistency condition:
	if $\varepsilon=10^{-2}$ the zone plotted in light grey between the dotted colored lines and the bound on $\xi$ is admissible.}
	\label{fig:rs-q1-alpha=-1o4}
\end{minipage}\hfill
\begin{minipage}{.49\columnwidth}
	\centering
	\includegraphics[width=1.01\textwidth]{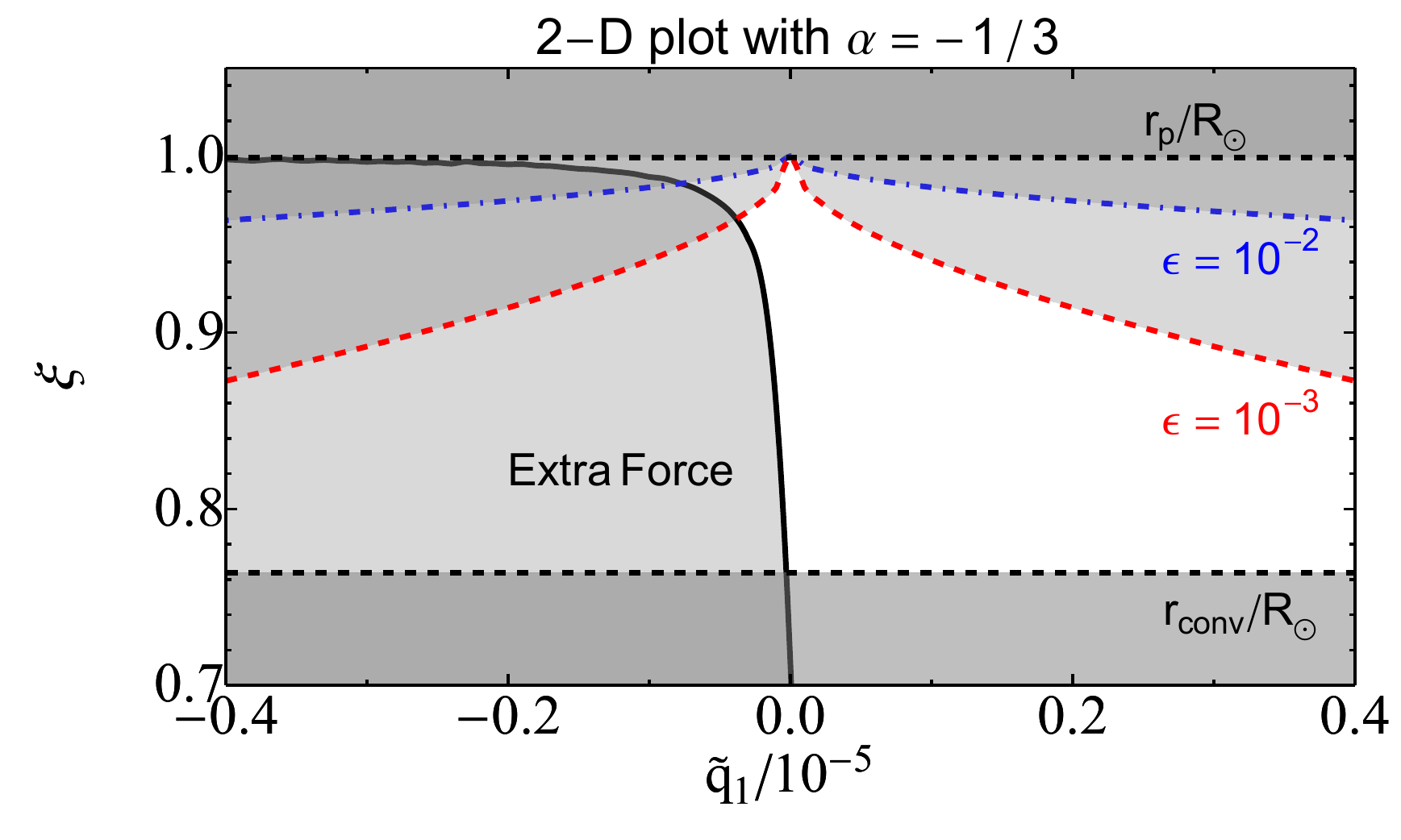}
	\caption{Constraint from extra force on the half plane $\widetilde q_1,\xi$ for $\alpha=-1/3$. The lower bound on the screening radius increases, but it decreases
	if $\alpha$ approaches the value $\alpha=-1$. The dotted colored lines yield the bound from the consistency condition: 
	if $\varepsilon=10^{-2}$ the zone plotted in light grey between the dotted colored lines and the bound on $\xi$ is admissible.}
	\label{fig:rs-q1-alpha=-1o3}
\end{minipage}
\end{figure}
\end{widetext}

\begin{widetext}
	
\begin{figure}[H]
\centering
\begin{minipage}{.49\columnwidth}
	\centering
	\includegraphics[width=1.01\textwidth]{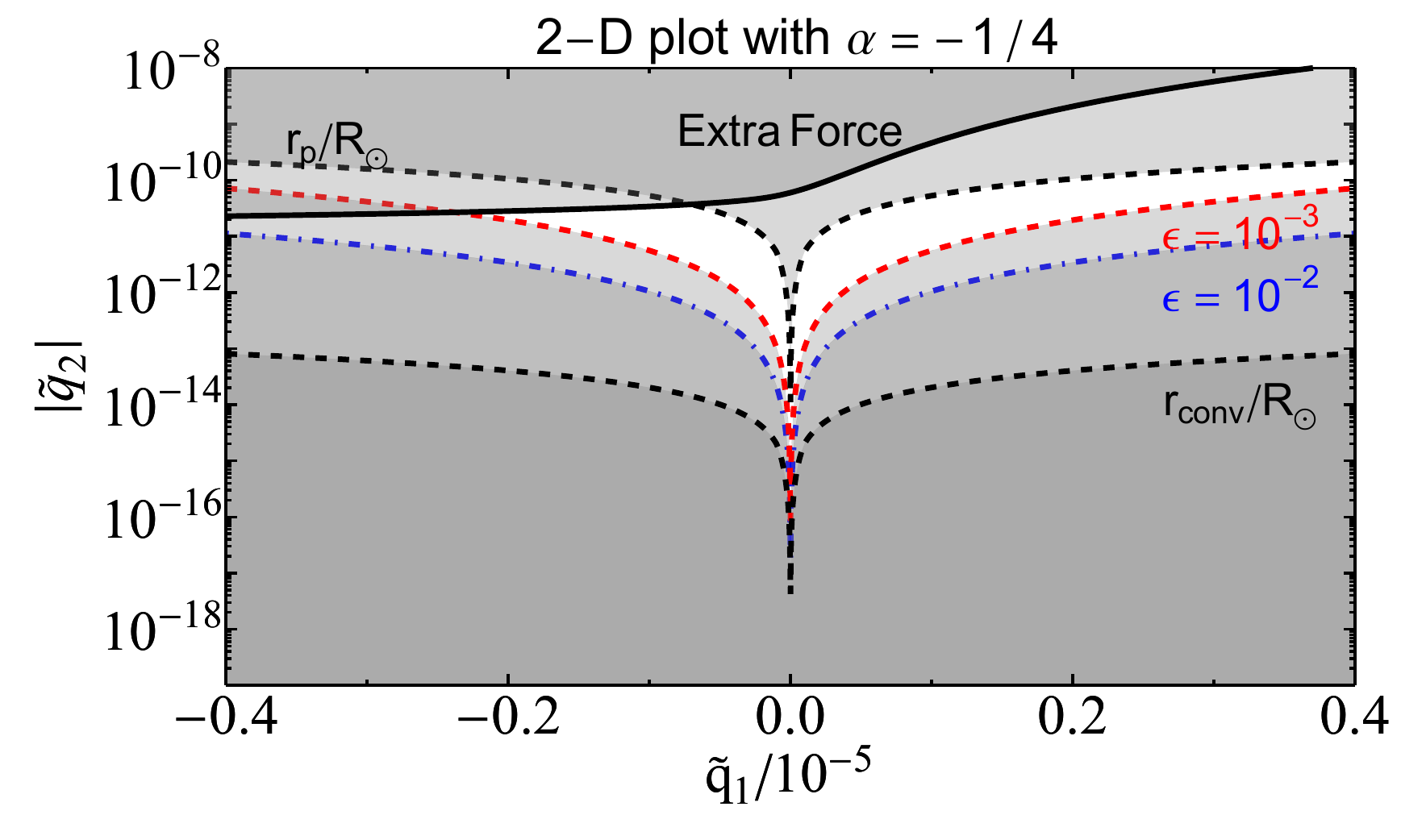}
	\caption{Constraint from extra force on the half plane $\widetilde q_1,|\widetilde q_2|$ for $\alpha=-1/4$. The solid line yields the upper bound on $|\widetilde q_2|$
	from spectroscopic measurements. The dotted colored lines yield the bound from the consistency condition:
	if $\varepsilon=10^{-2}$ the zone plotted in light grey between the dotted colored lines and the bound on $|\widetilde q_2|$ is admissible.}
	\label{fig:q1-q2-alpha=-1o4}
\end{minipage}\hfill
\begin{minipage}{.49\columnwidth}
	\centering
	\includegraphics[width=1.01\textwidth]{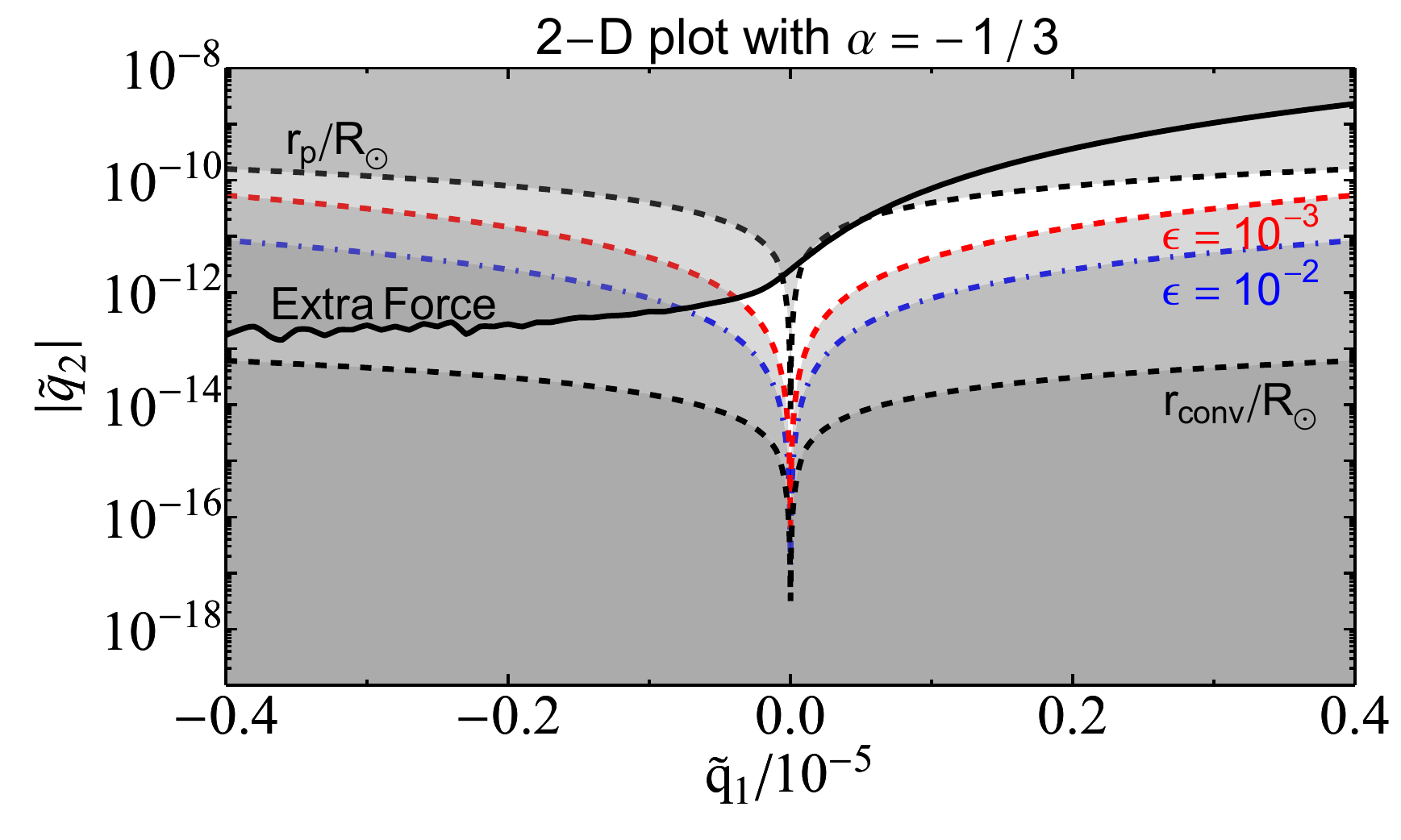}
	\caption{Constraint from extra force on the half plane $\widetilde q_1,|\widetilde q_2|$ for $\alpha=-1/3$. The upper bound on $|\widetilde q_2|$ decreases, but it increases
	if $\alpha$ approaches the value $\alpha=-1$. The dotted colored lines yield the bound from the consistency condition: 
	if $\varepsilon=10^{-2}$ the zone plotted in light grey between the dotted colored lines and the bound on $|\widetilde q_2|$ is admissible.}
	\label{fig:q1-q2-alpha=-1o3}
\end{minipage}
\end{figure}
\end{widetext}

The numerical computation shows that the constraint from spectroscopic measurements imposes a lower bound on the screening radius which lies in
the convection zone for $-1<\alpha<0$. Conversely, for $\alpha\leq-1$ the screening radius lies in Sun's radiative interior, so that for such values of $\alpha$
the constraint from the extra force is surely much weaker than the constraint from Cassini measurement.

Our results are graphically reported in Figures \ref{fig:rs-q1-alpha=-1o4}-\ref{fig:q1-q2-alpha=-1o3}: admissible regions for parameters are plotted in white,
while the excluded regions are plotted in grey. We denote $r_{\rm conv}$ the radius at the base of the convection zone.
Fig. \ref{fig:rs-q1-alpha=-1o4} shows the admissible region in the half plane $(\widetilde q_1,\xi)$ for $\alpha=-1/4$
and for values $\varepsilon=10^{-2}$ and $\varepsilon=10^{-3}$, while Fig. \ref{fig:rs-q1-alpha=-1o3} shows the admissible region for $\alpha=-1/3$.

Figures \ref{fig:q1-q2-alpha=-1o4} and \ref{fig:q1-q2-alpha=-1o3} show the admissible region in the half plane with coordinates $(\widetilde q_1,|\widetilde q_2|)$ for
$\alpha=-1/4$ and $\alpha=-1/3$, respectively. The value $\alpha=-1/3$ has been used in Ref. \cite{drkmattgal} to model the rotation curves of galaxies.
Figure \ref{fig:q1-q2-alpha=-1o3} shows some numerical noise on the left side of the solid curve which marks the upper bound on $|\widetilde q_2|$ from the extra force,
nevertheless such a noise does not affect importantly the behavior of the bound, moreover it lies in the excluded region.

When the screening radius lies deeply in the convection zone the Taylor approximation of the effective mass and of the integral in the equation (\ref{rs-integr-equation})
cannot be used, then the integrals are evaluated analytically by approximating the polytropic index in the convection zone with $n_c\approx 2$
(see Section \ref{sec:convection} of Appendix B).

Eventually, we argue that the constraint from extra force becomes competitive with the constraint from Cassini measurement for suitable values of $\alpha$ in the range
$-1<\alpha<0$. Further improvements could be expected from new spectroscopic measurements in solar missions like ESA Solar Orbiter \cite{Muller-Cyr}
and NASA Parker Solar Probe \cite{Fox-Velli}.

\section{Conclusions}

In this work we obtain Solar System bounds on an alternative theory of gravity with nonminimal coupling between curvature and matter, that has been introduced in Ref. \cite{BBHL},
by suitably implementing a screening mechanism which is the NMC version of the well known chameleon mechanism \cite{KW,HS}.
An analytic approximate solution of the field equations that exhibits the screening mechanism has been computed inside and around the Sun.
Such a solution shows the typical features of a chameleon solution, though significantly affected by the nonminimal coupling: i) a screening radius $r_s$ in the Sun's interior
that has to be determined, and such that GR is approximately satisfied for $r<r_s$ if a suitable consistency condition on the parameters of the model is also satisfied; 
ii) a thin shell close to the Sun's edge where the solution interpolates between the values in the Sun's interior and the values in the outskirts of the Solar System;
iii) the existence of the thin shell permits to satisfy stringent constraints from Solar System gravity experiments.

A specific feature of the nonminimal coupling between geometry and matter is the presence of a non-Newtonian extra force which, together with a fifth force
of the Yukawa type, appears in the solar interior and atmosphere modeled as a perfect fluid.
Constraints on the parameters of the gravity model have been computed from the Cassini measurement of PPN parameter $\gamma$,
and from the effect in the solar atmosphere of the extra force by resorting to spectroscopic measurements.

Our results show that taking into account all known solar physics bounds, the Cassini measurement of the parameter $\gamma$ constrains the parameters for the specific model Eq. (\ref{f1-f2-specific}), under conditions Eq. (\ref{q2-sign}) and rescaled according to Eq. (\ref{q1q2-rescaled}),
to be order $|\widetilde q_2| < 10^{-12}$ for $|\widetilde q_1| < 3 \times 10^{-6}$ and $-10 < \alpha <  -1$. 
These are specific constraints obtained under the assumption that the galaxy itself is screened within a distance from its center corresponding to the location of
the Solar System. The assessment of such an assumption requires an extension of the Solar System to galaxy analysis carried out in the present paper. Particularly,
one should look for a chameleon solution of the field equations in the Milky Way considering the
transition from the outskirts of the galactic halo to the large-scale structure of the Universe (see Ref. \cite{HS} for the analogous problem in $f(R)$ gravity).
Such an analysis will impose additional constraints on the parameters of the NMC gravity model that have to be compared with the Solar System bounds obtained in the
present paper. Moreover, it could also help to better understand the role of different powers $\alpha$ of curvature in the specific model Eq. (\ref{f1-f2-specific}) at different scales, as it
has been conjectured in Ref. \cite{drkmattgal}.
Eventually, the extension of the analysis in the present paper to the gravitational interaction between the Sun, planets and satellites should also provide further constraints
on the parameters of the gravity model. All these issues will be investigated in future research.

\appendix
\section*{Appendix A}

We give a proof a posteriori of inequalities Eqs. (\ref{deriv-potential-condit}) and (\ref{q1-ll-bound}) that have been used in order to compute the solution $\eta$ of equation (\ref{trace-approx}).
Hence, we prove that the solution (\ref{eta-solution-complete}) found for function $\eta$ is consistent with such inequalities.

\subsection{Verification of inequality $|\partial V/ \partial\eta|\ll \rho c^2/3$}\label{sec:ineq-V-rho}

The inequality is equivalent to inequality (\ref{omega-condition}) which we then consider.
We prove that for $r>r_s$ and $r$ close to $r_s$, the solution (\ref{eta-solution-complete}) for $\eta$ is such that curvature $R=\omega(\eta,\rho)$ 
quickly becomes much smaller than the GR curvature, so that inequality (\ref{omega-condition}) is verified a posteriori.

We use formula (\ref{R-omega-formula}) for curvature which we write in the form
\begin{equation}
\frac{c^2}{8\pi G}\frac{\omega(\eta,\rho)}{\rho} = \left(\frac{\rho}{\rho_s}\right)^{\alpha/(1-\alpha)}\left(\frac{N}{D}\right)^{1/(1-\alpha)}.
\end{equation}
Using Eqs. (\ref{lambda-expression}) and (\ref{R-omega-formula}) we have
\begin{equation}
N = \frac{\lambda_s^2}{6(1-\alpha)}\rho_s, \qquad D = \frac{c^2}{16\pi G}-q_1\rho-\frac{1}{2c^2}\eta.
\end{equation}
We approximate the denominator $D$ by means of a second order Taylor expansion around $r_s$, so that,
using the expression (\ref{eta-solution-complete}) for $\eta$, for $\rho(r)\gg\rho_g$ we have
\begin{eqnarray}
& &-\frac{1}{24\pi}\frac{M_{\rm eff}(r)}{r} + \frac{1}{6}\int_{r_s}^r\rho(r^\prime)r^\prime dr^\prime \nonumber\\
&=& \frac{1}{12} r^2 \rho_s\left(1-\frac{r_s}{r}\right)^2 + O\left(\left(1-\frac{r_s}{r}\right)^3\right),
\end{eqnarray}
and
\begin{equation}
D \approx \frac{\lambda_s^2}{6(1-\alpha)}\left[\rho_s + \alpha\rho_s^\prime r_s\left(1-\frac{r_s}{r}\right)\right]
+ A_s\left(1-\frac{r_s}{r}\right)^2,
\end{equation}
where
\begin{equation}
A_s = \frac{r_s^2}{2}\left[ \frac{1}{6}\rho_s - q_1\nabla^2\rho(r_s) \right].
\end{equation}
Since density $\rho(r)$ is decreasing, we have $\alpha\rho_s^\prime>0$ for $\alpha<0$ from which, using
the consistency condition (\ref{consist-condition}) for small $\lambda_s$ in the form (\ref{consist-cond-10-3}), a lower bound for $D$ follows for $\varepsilon<1/6$ and $r>r_s$:
\begin{equation}
D > \frac{1}{12}\rho_s r_s^2(1-6\varepsilon)\left(1-\frac{r_s}{r}\right)^2.
\end{equation}
Then, using this lower bound and Eq. (\ref{lambda-rescaled-lambdag}), we obtain an upper bound on the ratio $\omega/\rho$:
\begin{widetext}
\begin{equation}\label{omega-over-rho-upperb}
\frac{c^2}{8\pi G}\frac{\omega(\eta,\rho)}{\rho} < \frac{\rho_g}{\rho_s} \left(\frac{\rho}{\rho_s}\right)^{\alpha/(1-\alpha)}
\left[ \frac{2}{(1-\alpha)(1-6\varepsilon)}\left(\frac{\lambda_g}{r_s}\right)^2\right]^{1/(1-\alpha)}
\left(1-\frac{r_s}{r}\right)^{2/(\alpha-1)}.
\end{equation}
\end{widetext}
Now we observe that, for given $\alpha$, $q_1$ and $r_s$, the Yukawa range $\lambda_g$ is determined by formula (\ref{lambda-expression})
and by the integral equation (\ref{rs-integr-equation}):
\begin{eqnarray}\label{lambda_g-from-integral}
& &\frac{1}{6}\int_{r_s}^{r_g}\rho\left[ 1 - \left(\frac{\rho_g}{\rho}\right)^{-\alpha\slash(1-\alpha)} \right]r dr \\
&=& -q_1(\rho_s+\rho_s^\prime r_s) + \frac{\lambda_g^2}{6(1-\alpha)}\rho_g. \nonumber
\end{eqnarray}
If we compute $\lambda_g$ from Eq. (\ref{lambda_g-from-integral}), using approximation (\ref{second-integral}) and inequality (\ref{zconv-second-ineq-eps}),
the effect of the term with $q_1$ is a fraction of the term with $\lambda_g$ for $\varepsilon<10^{-2}$ and starts to become negligible for $\varepsilon<10^{-3}$.
Since we are interested in order of magnitude estimates, we neglect the term with $q_1$ in the following computations. An analogous result
holds if $r_s$ lies in the solar atmosphere.

Let us give two examples that show for which values of $r$ the desired inequality (\ref{omega-condition}) is achieved for given $\alpha$ and $r_s$ in the convection zone.
By these examples one can understand that the results obtained by using inequality (\ref{omega-over-rho-upperb}) are quite general.

Let us consider the value $r_s\approx 0.98R_\odot$ for the screening radius which,
using the model of density profile in the convection zone reported in Appendix B, corresponds to $\rho_s\approx 3\times 10^{-4}\,{\rm g\,cm^{-3}}$,
and makes the constraint from Cassini measurement satisfied for $-10\leq\alpha\leq -1$ and $\varepsilon\leq 10^{-2}$.

We fix $\varepsilon=10^{-2}$ since the result does not depend significantly on $\varepsilon$ if $\varepsilon$ is small enough (see inequality (\ref{omega-over-rho-upperb})), 
and we first set $\alpha=-1$ which implies, using Eq. (\ref{lambda_g-from-integral}), $\lambda_g \approx 1.87\times 10^9 R_\odot \approx 8.7\times 10^6$ AU.
In this case inequality (\ref{omega-over-rho-upperb}) becomes
\begin{eqnarray}
\frac{c^2}{8\pi G}\frac{\omega(\eta,\rho)}{\rho} &<& \frac{\rho_g}{\sqrt{(1-6\varepsilon)\rho_s\rho}} \, \frac{\lambda_g}{r_s}\left(1-\frac{r_s}{r}\right)^{-1} \nonumber\\
&<& \frac{2\times 10^{-10}}{1-r_s\slash r} \,,
\end{eqnarray}
where for the last upper bound we have used $\rho(r)>\rho(r_p)\approx 2.73\times 10^{-7}\,{\rm g\,cm^{-3}}$, where $r_p$ is the radius
at the top of the convection zone. Inequality (\ref{omega-condition}) then follows:
\begin{equation}
\frac{c^2}{8\pi G} \omega(\eta,\rho) < 10^{-3}\rho \quad\mbox{for } r-r_s>2\times 10^{-7}r_s \approx 140\,{\rm m},
\end{equation}
with $10^{-7}r_s\ll R_\odot-r_s\approx 13900\,{\rm km}$. Hence, for $r>r_s$ and $r$ very close to $r_s$ there is a sharp transition to a low value of curvature
(much smaller than GR curvature) which is a typical property of a chameleon solution (see also Refs. \cite{KW,HS}).

If we now set $\alpha=-2$ we get $\lambda_g\approx 2.29\times 10^9 R_\odot \approx 1.1\times 10^7$ AU, and
\begin{eqnarray}
\frac{c^2}{8\pi G}\frac{\omega(\eta,\rho)}{\rho} &<& \rho_g \left[\frac{2}{3(1-6\varepsilon)\rho_s}\right]^{1/3}
\left[\frac{\lambda_g}{r_s\rho\left( 1-r_s\slash r \right)}\right]^{2/3} \nonumber\\
&<& \frac{10^{-12}}{\left(1-r_s\slash r\right)^{2/3}},
\end{eqnarray}
from which inequality (\ref{omega-condition}) follows:
\begin{equation}
\frac{c^2}{8\pi G} \omega(\eta,\rho) < 10^{-6}\rho \quad\mbox{for } r-r_s>10^{-9}r_s \approx 70\,{\rm cm}.
\end{equation}
Increasing $|\alpha|$ the transition to low curvature becomes sharper.

\subsection{Verification of inequality $|q_1| \ll |q_2| R^{\alpha-1}$}\label{sec:q1-q2-Rexp-alpha}

Using Eq. (\ref{R-omega-formula}) with $R=\omega(\eta,\rho)$, the inequality is written in the form
\begin{equation}\label{q1-q2-equiv-ineq}
\frac{16\pi G}{c^2} |\alpha q_1| \rho \ll 1 - \frac{8\pi G}{c^4}\eta - \frac{16\pi G}{c^2}q_1\rho.
\end{equation}
Using now expression (\ref{eta-solution-complete}) for $\eta$, neglecting $\lambda_s$ and assuming the screening radius in the solar convection zone, 
we approximate $M_{\rm eff}(r)\approx M_{\rm eff}(R_\odot)$ and we get
\begin{widetext}
\begin{equation}
1-\frac{8\pi G}{c^4}\eta \approx \frac{16\pi G}{c^2}q_1\rho_s + \frac{16\pi G}{c^2}q_1\rho_s^\prime r_s\left(1-\frac{r_s}{r}\right)
-\frac{2G}{3c^2}\frac{M_{\rm eff}(R_\odot)}{r} + \frac{8\pi G}{3c^2}\int_{r_s}^{R_\odot}\rho(r)rdr,
\end{equation}
\end{widetext}
from which, using the inequality
\begin{equation}
4\pi \int_{r_s}^{R_\odot}\rho(r)rdr > \frac{4\pi}{R_\odot}\int_{r_s}^{R_\odot}\rho(r)r^2dr \approx \frac{M_{\rm eff}(R_\odot)}{R_\odot},
\end{equation}
we obtain
\begin{eqnarray}
1-\frac{8\pi G}{c^4}\eta &>& \frac{16\pi G}{c^2}q_1\rho_s + \frac{16\pi G}{c^2}q_1\rho_s^\prime r_s\left(1-\frac{r_s}{r}\right) \nonumber\\
&+& \frac{2G}{3c^2}\,\frac{M_{\rm eff}(R_\odot)}{R_\odot}\left(1-\frac{R_\odot}{r}\right).
\end{eqnarray}
Using now inequality (\ref{zconv-first-ineq-eps}), valid for $r_s$ in the convection zone, we find the following lower bounds:
\begin{widetext}
\begin{eqnarray}
1-\frac{8\pi G}{c^4}\eta &>& \frac{16\pi G}{c^2}q_1\rho_s + \frac{2G}{3c^2}\,\frac{M_{\rm eff}(R_\odot)}{r_s}
\left[ (1-20\varepsilon) \left(1-\frac{r_s}{r}\right) -\left(1-\frac{r_s}{R_\odot}\right) \right], \qquad\mbox{for }q_1>0, \nonumber\\
1-\frac{8\pi G}{c^4}\eta &>& \frac{16\pi G}{c^2}q_1\left\{ \rho_s+\rho_s^\prime r_s \left[ 1-\frac{r_s}{r}\left(1+\frac{1}{20\varepsilon}
\left(1-\frac{r}{R_\odot}\right)\right)\right]\right\}, \qquad\mbox{for }q_1<0. \label{lower-bound-q1-neg}
\end{eqnarray}
\end{widetext}
Let us first consider the case $q_1>0$. In this case the lower bound can be written in the form
\begin{equation}
1-\frac{8\pi G}{c^4}\eta > \frac{16\pi G}{c^2}q_1\rho_s + \phi(r,r_s,\varepsilon),
\end{equation}
where the function $\phi(r,r_s,\varepsilon)$ is positive if the following condition is satisfied:
\begin{equation}\label{q1-suff-cond}
r > \left[ 1 + \delta(\varepsilon,r_s) \right] R_\odot, 
\end{equation}
with
\begin{equation}
\delta(\varepsilon,r_s) = \frac{20\varepsilon(1-r_s\slash R_\odot)}{r_s\slash R_\odot-20\varepsilon}.
\end{equation}
Let now $r$ in the solar atmosphere, $r_s$ in the convection zone, and $\varepsilon>0$ be such that $\phi(r,r_s,\varepsilon)$ is positive and
$\rho(r)\ll \rho_s=\rho(r_s)$. Then we have
\begin{equation}\label{|alpha|-equiv-ineq}
\frac{16\pi G}{c^2}q_1\rho(r) \ll \frac{16\pi G}{c^2}q_1\rho_s < 1-\frac{8\pi G}{c^4}\eta,
\end{equation}
so that inequality (\ref{q1-q2-equiv-ineq}) is satisfied if the condition $|\alpha|\rho(r)\ll\rho_s$ is also satisfied.

For $r_s<r_p$, where $r_p=R_\odot-500\,{\rm km}$ is the radius at the base of the photosphere (see Appendix B), the function $\delta(\varepsilon,r_s)$ is
decreasing as $r_s$ increases and increasing as $\varepsilon$ increases.
Particularly, for $\varepsilon$ small enough, in order to have inequality (\ref{q1-q2-equiv-ineq}) satisfied for
$r$ varying in the inner regions of the solar atmosphere, $r_s$ has to be close enough to $R_\odot$,
which means that the contribution to the effective mass only comes from a thin shell in the upper part of the solar interior, but this is just
the requirement imposed by the constraint from the Cassini measurement (see Section \ref{sec:Cassini-measure}).

Let us give some examples that show for which values of $r,r_s,\varepsilon,\alpha$, with $r_s$ satisfying the Cassini constraint,
the desired inequality (\ref{q1-q2-equiv-ineq}) is satisfied. Let us consider the value $r_s=0.98R_\odot$ which
corresponds to $\rho_s\approx 3\times 10^{-4}\,{\rm g\,cm^{-3}}$. 

For $\varepsilon=10^{-2}$ we have $\delta=0.005$
so that, using the density model of the solar atmosphere reported in Appendix B, $r_0=(1+\delta)R_\odot$ is located in the chromosphere-corona
transition region, and for $r>r_0$ we have $\rho(r)<\rho(r_{\rm cr})=9.82\times 10^{-15}\,{\rm g\,cm^{-3}}$ (density at the top of the chromosphere). 
Then $|\alpha|\rho(r)\ll \rho_s$ for $|\alpha|<10^8$ so that for such values of $\alpha$ inequality (\ref{q1-q2-equiv-ineq}) is satisfied.

For $\varepsilon=10^{-3}$ we have $\delta=4\times 10^{-4}$, with $r_0=(1+\delta)R_\odot$ located in the chromosphere so that, using the density profile of the
chromosphere reported in Appendix B, for $r>r_0$ we have $\rho(r)<\rho(1.0004R_\odot)\approx 10^{-9}\,{\rm g\,cm^{-3}}$.
In this case inequality (\ref{q1-q2-equiv-ineq}) is satisfied for $|\alpha|<10^3$.

If now $r_s$ increases and reaches the top of the convection zone, $r_s=r_p$, then $\rho_s=2.73\times 10^{-7}\,{\rm g\,cm^{-3}}$ and the Cassini constraint
is satisfied. For $\varepsilon=10^{-2}$ we have $\delta=1.8\times 10^{-4}$, and now we need to take values of $r$ larger than the ones implied by the inequality
$r>(1+\delta)R_\odot$. If we take $r_0=1.001R_\odot$ in the chromosphere, then for $r>r_0$ we have $\rho(r)<6.99\times 10^{-11}\,{\rm g\,cm^{-3}}$, 
and inequality (\ref{q1-q2-equiv-ineq}) is satisfied for $|\alpha|<39$.

We consider now the case $q_1<0$. We need the expression of $\rho_s^\prime r_s$ which is obtained by computing the derivative of the density profile
in the convection zone given by Eqs. (\ref{conv-density-prof}) and (\ref{conv-temp-prof}) in Appendix B:
\begin{equation}
\rho_s^\prime r_s = -n_c \frac{GM_\odot}{C_p}\frac{\rho_s}{GM_\odot\slash C_p - T_0 r_s},
\end{equation}
where $n_c$ is an effective polytropic index, $C_p$ is an averaged value of the specific heat at constant pressure, and $T_0$ is a reference temperature
(see Appendix B for details and numerical values). Then the lower bound (\ref{lower-bound-q1-neg}) can be written in the form
\begin{equation}
1-\frac{8\pi G}{c^4}\eta > \frac{16\pi G}{c^2} \left\vert q_1 \right\vert \rho_s \left[ \psi(r,r_s,\varepsilon) - 1 \right],
\end{equation}
where
\begin{widetext}
\begin{equation}
\psi(r,r_s,\varepsilon) = n_c \frac{GM_\odot\slash C_p}{GM_\odot\slash C_p - T_0 r_s}
\left[ 1-\frac{r_s}{r}\left(1+\frac{1}{20\varepsilon} \left(1-\frac{r}{R_\odot}\right)\right)\right].
\end{equation}
\end{widetext}
The function $\psi(r,r_s,\varepsilon)$ is positive in the domain $D=\{r\geq R_\odot,0\leq r_s \leq r_p,\varepsilon\geq 0\}$, and in this domain it achieves
the absolute minimum, independent of $\varepsilon$, at $r=R_\odot$, $r_s=r_p$ with value ${\rm min}_D\psi(r,r_s,\varepsilon)\approx 1.918$.
Let now $r$ in the solar atmosphere and $r_s$ in the convection zone be such that
$|\alpha|\rho(r)\ll 0.918 \rho_s$. Then we have
\begin{eqnarray}
& &\frac{16\pi G}{c^2}\left\vert \alpha q_1 \right\vert \rho(r) \ll 0.918 \frac{16\pi G}{c^2}\left\vert q_1\right\vert \rho_s \nonumber\\
&<& \frac{16\pi G}{c^2}\left\vert q_1\right\vert \rho_s\left[ \psi(r,r_s,\varepsilon) - 1 \right] < 1-\frac{8\pi G}{c^4}\eta,
\end{eqnarray}
so that, being $q_1<0$, inequality (\ref{q1-q2-equiv-ineq}) is satisfied for any $\varepsilon>0$.

We now give two examples. Let $r$ and $r_s$ be such that $r>R_\odot$ and $r_s=0.98 R_\odot$. Then, using the density profile in the chromosphere,
we have $\rho(r)<5.9\times 10^{-9}\,{\rm g\,cm^{-3}}$ (density at the bottom of the chromosphere), and inequality (\ref{q1-q2-equiv-ineq}) is satisfied for
$|\alpha|<10^2$. If now $r$ and $r_s$ are such that $r>r_0$, with $r_0=1.001R_\odot$ in the chromosphere, and $r_s=r_p$ at the top of the convection zone, 
then inequality (\ref{q1-q2-equiv-ineq}) is satisfied for $|\alpha|<35$.

Analogous results can be obtained if the screening radius is located in the photosphere or outside of the photosphere.

\appendix
\section*{Appendix B}

We report on a model of solar mass density profile $\rho(r)$, both for the interior and the atmosphere, which is
used in order to find analytical estimates of the constraints on the NMC gravity parameters at a suitable order of magnitude.
Matter in the Sun is modeled as a perfect gas in hydrostatic equilibrium, with the exception of the outer corona,
where dynamical equilibrium of a stationary atmosphere is used.

\subsection{Solar interior}

The radius of the Sun $R_\odot=6.9634\times 10^5\,{\rm km}$ is defined to be the radius of the edge, or limb, of the Sun when observed in white light.
The solar atmosphere begins below the spherical surface of radius $R_\odot$ and center in the origin \cite{Tan-Han}, 
at a depth of about 500 km, and extends outward from the Sun. Then $r_p=R_\odot-500\,{\rm km}$ is the radius at the base of the photosphere. 
The density profile is computed for Newtonian gravity, while
a computation based on the solution of the Lane-Emden equation in the presence of an Yukawa force can be found in Ref. \cite{laneemdenyukawa}.

\subsubsection{Convection zone}\label{sec:convection}

This outer zone of the solar interior is important since the screening radius that saturates the Cassini bound (\ref{Cassini-upper-bound})
lies in such a zone. We use a polytrope model with an effective polytropic index $n_c=2.33$ \cite{Mullan}:
\begin{equation}\label{conv-density-prof}
\rho(r) = K_c\left[ T(r) \right]^{n_c},
\end{equation}
with $K_c=3.44\times 10^{-16}$, and the radius $r$ varying in the range
\begin{equation}
r_{\rm conv} \leq r < r_p, \qquad r_{\rm conv} = 5.3185\times 10^5 \,\mbox{km}.
\end{equation}
The temperature profile is approximated by \cite{Mullan}
\begin{equation}\label{conv-temp-prof}
T(r) = \frac{GM_\odot}{C_p r} - T_o,
\end{equation}
with $M_\odot=1.989\times 10^{33}\,{\rm g}$,
$C_p=2.95\times 10^8 \,{\rm erg\,g^{-1}\,K^{-1}}$ is an averaged value of the specific heat at constant pressure,
and $T_o=6.461\times 10^6 \,{\rm K}$.

The density values computed by means of the above model
and by means of solar models that incorporate helioseismological measurements \cite{Bahcall} are of the same order of magnitude.

In order to verify the consistency condition (\ref{consist-decomposed}) the expression of the Laplacian of density is required:
\begin{equation}
\nabla^2\rho(r) = n_c(n_c-1)\left(\frac{GM_\odot}{C_p}\right)^2\frac{K_c}{r^4}\left[T(r\right)]^{n_c-2}.
\end{equation}
We have $\nabla^2\rho(r)>0$ in the convection zone and $\nabla^2\rho\slash\rho$ is an increasing function,
so that the maximum of the quantity $|\nabla^2\rho|\slash\rho$ is achieved at $r=r_p$ and it is of order of $10^7\slash R_\odot^2$.

\subsubsection{Radiative interior}\label{sec:radiative}

The Cassini bound prevents the screening radius from penetrating inside the radiative interior, hence the density profile
in this region will only be used for verifying the consistency condition (\ref{consist-condition}). We use the polytrope model with $n_r=3$:
\begin{equation}
\rho(r) = K_r\left[ T_6(r) \right]^{n_r}, \qquad r < r_{\rm conv},
\end{equation}
with $K_r=2.6885\times 10^{-2}$ and $T=10^6 T_6 \,{\rm K}$. The temperature profile is estimated by resorting to an approximation
of the gravitational acceleration profile inside the zone (see Ch. 9 of Ref. \cite{Mullan} for details):
\begin{equation}
T(r) = T(r_{\rm conv}) + T_\ast \left[ \chi(r)-\frac{R_\odot}{r_{\rm conv}} \right],
\end{equation}
where
\begin{equation}
T_\ast = \frac{\mu m_p}{(n_r+1)k_B}\,\frac{GM_\odot}{R_\odot},
\end{equation}
and
\begin{equation}
\chi(r) =
\begin{cases}
2\left[ 3-\left(4r\slash R_\odot\right)^2 \right] & 0\leq r < R_\odot\slash 4, \\
R_\odot\slash r & R_\odot\slash 4 \leq r < r_{\rm conv},
\end{cases}
\end{equation}
$T(r_{\rm conv})=2\times 10^6\,{\rm K}$, $m_p=1.66\times 10^{-24}\,{\rm g}$ is the proton mass, $\mu=0.62$ is the mean molecular weight,
and $k_B=1.3806\times 10^{-16}\,{\rm erg \, K^{-1}}$
is the Boltzmann constant. This density profile is used up to $r=0$.

Since the temperature profile is continuous across the base $r=r_{\rm conv}$ of the convection zone and the polytropic index changes
from the convection zone to the radiative interior, the density model introduces an artificial discontinuity at $r=r_{\rm conv}$. Since the density
profile in the inner region of the convection zone and in the radiative interior is only used for verifying the consistency condition (\ref{consist-decomposed}),
which requires the computation of the Laplacian of mass density, then the consistency condition is verified in the separate regions without crossing the
discontinuity.

For $0\leq r < R_\odot\slash 4$ the Laplacian of density is given by 
\begin{eqnarray}
\nabla^2\rho &=& 6.4\times 10^{-11} n_r \frac{K_r T_\ast}{R_\odot^2} \left[T_6(r)\right]^{n_r-2} \nonumber\\
&\times& \left[ T_\ast \left( 224\frac{r^2}{R_\odot^2} -3\frac{R_\odot}{r_{\rm conv}} -18 \right)-3T(r_{\rm conv}) \right], \nonumber\\
\end{eqnarray}
and for $R_\odot\slash 4 \leq r < r_{\rm conv}$ it is given by
\begin{equation}
\nabla^2\rho = 10^{-12}n_r(n_r-1)K_r T_\ast^2 \left[ T_6(r) \right]^{n_r-2}\frac{R_\odot^2}{r^4}.
\end{equation}
Using the above expressions we find that the maximum of the quantity $|\nabla^2\rho|\slash\rho$ in the radiative interior
is achieved at $r\approx 0.22 R_\odot$ and it is of order of $10^2\slash R_\odot^2$.

\subsection{Solar atmosphere}\label{sec:solar-atmosph}

Large density gradients take place in the solar atmosphere, particularly in the chromosphere-corona transition region.
Then the extra force, which depends on the density gradient, 
can become a significant perturbation of hydrostatic equilibrium in such regions with large gradients.

\subsubsection{Photosphere}\label{sec:photosphere}

For the density in the photosphere we use the following model, adapted from \cite{DeJager}, for $r_p\leq r < R_\odot$:
\begin{equation}
\rho(r) = \frac{\mu m_p p_m}{k_B[T_m+A(R_\odot-r)^2])}\exp\left(\frac{R_\odot-r}{H_p}\right),
\end{equation}
where $\mu=1.26$,
$H_p=117\,{\rm km}$, $A=8.8\times 10^{-3}\,{\rm km^{-2}K}$, $T_m=4.4\times 10^3\,{\rm K}$ is the temperature minimum at the top of the photosphere,
$p_m$ is pressure corresponding at the temperature minimum, such that $p_m\slash k_B=1.2\times 10^{19}\,{\rm K\,cm^{-3}}$.

\subsubsection{Chromosphere}\label{sec:chromosph}

The middle chromosphere is characterized by a broad temperature plateau at $T\approx 6500$ K \cite{Vernazza}, hence, for simplicity, we 
adopt an isothermal model for this layer of the solar atmosphere:
\begin{equation}\label{chromo-dens-profile}
\rho(r) = \rho_p\exp\left( -\frac{r-R_\odot}{H_c} \right), \quad R_\odot \leq r < 1.003 R_\odot,
\end{equation}
with 
\begin{equation}
H_c = \frac{k_B TR_\odot^2}{\mu m_pGM_\odot}\approx 157 \,{\rm km}, 
\end{equation}
$\mu=1.26$, and $\rho_p=5.9\times 10^{-9}\,{\rm g\,cm^{-3}}$ is density at the top of the photosphere
(value interpolated from the average quiet Sun model in \cite{Vernazza}).
The density profile (\ref{chromo-dens-profile}) coincides with the best fitting formula used in \cite{Kontar}.

\subsubsection{Chromosphere-corona transition region}\label{sec:chromo-coro-transr}

Here the steepest density gradient takes place. The density profile in this region is computed by using hydrostatic equilibrium together with the average emission measure
distribution given in Eq. (\ref{emission-measure}).
An analytic estimate of density is found by approximating the integral in Eq. (\ref{emission-measure}) as in \cite{Jordan76}, see Section \ref{sec:extra-f-atmosph}.
The resulting density profile $\rho\approx \mu m_p p_0\slash(k_BT)$ is given by
\begin{equation}
\rho(r) = \mu m_p \frac{p_0}{k_B} \left[ T_0^{b+2} + 0.231\frac{b+2}{\xi^2a}\left(\frac{p_0}{k_B}\right)^2(r-r_0) \right]^{-\frac{1}{b+2}},
\end{equation}
with $T_0=T(r_0)=1.5\times 10^5\,{\rm K}$, $p_0\slash k_B=2\times 10^{15}\,{\rm K\,cm^{-3}}$, and we use the values of $a,b$ reported in \cite{RayDoyle}
for the average quiet Sun. The transition region is divided into two zones:
 \begin{itemize}
\item[{\rm (i)}]
{\it zone with temperature $2\times 10^4\,{\rm K} \leq T<1.5\times10^5\,{\rm K}$}; in this zone we use $\mu=0.65$, $\xi=2$,
$b=-3.5$, $a=1\times 10^{43}$, and the zone corresponds to
\begin{equation}
0.003 R_\odot \approx 2089\,{\rm km} \leq r-R_\odot < 2186\,{\rm km};
\end{equation}
\item[{\rm (ii)}]
{\it zone with temperature $1.5\times10^5\,{\rm K} \leq T < 1\times 10^6\,{\rm K}$}; in this zone we use $\mu=0.62$, $\xi=1.91$,
$b=1.65$, $a=2\times 10^{16}$, and the zone corresponds to
\begin{equation}
2186\,{\rm km} \leq r-R_\odot < 3906\,{\rm km}.
\end{equation}
\end{itemize}
Density exhibits the steepest gradient in zone (i) which is denoted the lower transition region.
The thickness of the transition zone is in good agreement with the empirical atmospheric model computed in \cite{Mariska-book} (Ch. 6, Fig. 6.1)
using the emission measures tabulated in \cite{RayDoyle}.

\subsubsection{Inner corona}

We consider the inner corona approximately isothermal, so that the density profile is
\begin{equation}
\rho(r) = \rho_{\rm ic}\exp\left[ \frac{\mu m_pGM_\odot}{k_BT}\left(\frac{1}{r}-\frac{1}{r_{\rm ic}}\right) \right],
\end{equation}
with $\mu=0.62$,
\begin{equation}
r_{\rm ic} \leq r < r_a,\quad r_{\rm ic}=R_\odot+3906\,{\rm km}, \quad r_a=1.06 R_\odot,
\end{equation}
$r_a$ is the base of the outer corona, $\rho_{\rm ic}=2\times 10^{-15}\,{\rm g\,cm^{-3}}$ is density at the base of the inner corona, and
$T\approx 10^6\,{\rm K}$.

\subsubsection{Outer corona}\label{sec:outer-corona}

The model is based on the stationary expansion of the outer corona in which temperature $T$ is assumed constant
between the coronal base at $r=r_a$ and an outer boundary at $r=r_b$, with adiabatic expansion beyond $r_b$ \cite{Stix,Parker}.
The flow of the expanding atmosphere turns from subsonic to supersonic, giving rise to the solar wind, as it passes through a critical radius $r_c$
such that $r_a<r_c<r_b$.
Such a model yields values of solar wind speed and density which are in good agreement with measurements from the HELIOS solar probes \cite{Schwenn}
and Parker Solar Probe \cite{Allen}.

{\it Subsonic zone}. This zone corresponds to the range of distances $r_a\leq r < r_c$, where the critical radius $r_c$ is given by \cite{Stix,Parker}
\begin{equation}
r_c = \frac{\mu m_pGM_\odot}{2k_BT_a},
\end{equation}
where we choose $T_a=1\times 10^6\,{\rm K}$ for the constant value of temperature in the range $(r_a,r_b)$, and $\mu=0.54$ which corresponds to the average value
0.032 of the helium abundance \cite{Schwenn}, the ratio of helium to proton number density.
Then we get $r_c\approx 6.19R_\odot$.
If we set $\psi=v^2\slash c_s^2<1$, where $v=v(r)$ is the velocity of the expanding gas and $c_s$ is the isothermal speed of sound, in the region where
$\psi\ll\vert\ln(\psi)\vert$ the density profile is approximated by
\begin{equation}
\rho(r) = \rho_a \exp\left[2r_c\left(\frac{1}{r}-\frac{1}{r_a}\right)\right],
\end{equation}
where $\rho_a=9.7\times 10^{-16}\,{\rm g\,cm^{-3}}$ is density at the base of the outer corona. 

{\it Supersonic zone}. This zone corresponds to distances $r\geq r_c$ and it is further divided in two subzones.
In the first subzone the expansion is isothermal and it takes place in the range $r_c\leq r < r_b$.
We use the value $r_b=192R_\odot\approx 0.89$ AU which corresponds to a velocity $v(r_b)\approx 460\,{\rm km\,s^{-1}}$ of the
solar wind \cite{Parker}, in good agreement with observations \cite{Schwenn,Allen}. 

In the supersonic zone $\psi>1$, and at distances $r$ such that $\psi\gg\ln(\psi)$ the density profile in the first subzone is approximated by
\begin{equation}
\rho(r) = \rho_a \, \frac{\exp\left[3/2-2(r_c/r_a)\right]}{2\left[\ln(r/r_c)\right]^{1/2}} \left(\frac{r_c}{r}\right)^2,
\end{equation}
and we use this approximation in the range $(r_c,r_b)$.
In the second subzone $r\geq r_b$, the expansion is adiabatic, and density is approximated by
\begin{equation}\label{interplanet-density}
\rho(r) = \rho_a \, \frac{\exp\left[3/2-2(r_c/r_a)\right]}{2\left[\ln(r_b/r_c)\right]^{1/2}} \left(\frac{r_c}{r}\right)^2,
\end{equation}
hence density decreases as $1\slash r^2$.

\section*{Acknowledgments}

The work of R.M. is partially supported, and the work of M.M. and S.DA is fully supported, by 
INFN (Istituto Nazionale di Fisica Nucleare, Italy), as part of the MoonLIGHT-2 experiment in the framework 
of the research activities of the Commissione Scientifica Nazionale n. 2 (CSN2). 
C.G. is supported by the Fundo Regional para a Ciência e Tecnologia and Azores Government Grant No. M3.2DOCPROF/F/008/2020.




\begin{thebibliography}{99}
%
\bibitem{Capoz-1} 
S.~Capozziello, {\it Int. J. Mod. Phys. D} {\bf 11}, 483 (2002).
%
\bibitem{Carroll}
S.~M. Carroll, V.~Duvvuri, M.~Trodden and M.S.~Turner, {\it Phys. Rev. D} {\bf 70}, 043528 (2004).
%
\bibitem{Capoz-2}
S.~Capozziello and M.~De Laurentis, {\it Phys. Repts.} {\bf 509}, 167 (2011).
%
\bibitem{DeFTs}
A.~De~Felice and S.~Tsujikawa, {\it Liv. Rev. Rel.} {\bf 13}, 3 (2010).
%
\bibitem{BBHL}
O.~Bertolami, C. G.~B\"{o}hmer, T.~Harko and F. S. N.~Lobo,
Phys. Rev. D {\bf 75}, 104016 (2007).
%
\bibitem{drkmattgal}
O.~Bertolami and J.~P\'aramos, {\it JCAP} {\bf 009}, 1003 (2010).
%
\bibitem{curraccel}
O.~Bertolami, P.~Fraz\~ao and J.~P\'aramos, {\it Phys. Rev. D} {\bf 81}, 104046 (2010).
%
\bibitem{c}
C.~Gomes, J. G.~Rosa and O.~Bertolami, {\it JCAP} {\bf 06}, 021 (2017).
%
\bibitem{d}
O.~Bertolami, P.~Fraz\~ao and J.~P\'aramos, {\it JCAP} {\bf 05}, 029 (2013).
%
\bibitem{e}
O.~Bertolami, C.~Gomes and F. S. N.~Lobo, {\it Eur. Phys. J. C} {\bf 78}, 303 (2018).
%
\bibitem{f}
O.~Bertolami and C.~Gomes, {\it JCAP} {\bf 09}, 010 (2014).
%
\bibitem{g}
C.~Gomes, {\it Eur. Phys. J. C} {\bf 80}, 633 (2020).
%
\bibitem{h}
T. D.~Ferreira, N. A.~Silva, O.~Bertolami, C.~Gomes and A.~Guerreiro, {\it Phys. Rev. E} {\bf 101}, 023301 (2020).
%
\bibitem{i}
T.D.~Ferreira, J.~Novo, N.A.~Silva, A.~Guerreiro and O.~Bertolami, {\it Phys.Rev. D} {\bf 103}, 124019 (2021). 
%
\bibitem{MPBD}
R.~March, J.~P\'aramos, O.~Bertolami and S.~Dell'Agnello,
{\it Phys. Rev. D} {\bf 95}, 024017 (2017).
%
\bibitem{BLP}
O.~Bertolami, F.~Lobo and J.~P\'aramos,
{\it Phys. Rev. D} {\bf 78}, 064036 (2008).
%
\bibitem{MBMBD}
R.~March, O.~Bertolami, M.~Muccino, R.~Baptista and S.~Dell'Agnello, {\it Phys. Rev. D} {\bf 100},  042002 (2019).
%
\bibitem{Zum}
M.A.~Zumberge {\it et al}., {\it Phys. Rev. Lett.} {\bf 67}, 3051 (1991).
%
\bibitem{KW}
J.~Khoury and A.~Weltman,
{\it Phys. Rev. D} {\bf 69}, 044026 (2004).
%
\bibitem{HS}
W.~Hu and I.~Sawicki,
{\it Phys. Rev. D} {\bf 76}, 064004 (2007).
%
\bibitem{Cassini} B.~Bertotti, L.~Iess and P.~Tortora, {\it Nature} {\bf 425}, 374 (2003).
%
\bibitem{Fisher-Carlson} S.B. Fisher and E.D. Carlson, arXiv:2108.13492 [gr-qc].
%
\bibitem{multiscalar}O.~Bertolami and J.~P\'aramos, {\it Class. Quant. Grav.} {\bf 25}, 245017 (2008).
%
\bibitem{Sotiriou1}T. P.~Sotiriou and V.~Faraoni, {\it Class. Quant. Grav.} {\bf 25}, 5002 (2008).
%
\bibitem{stellequil}
O.~Bertolami and J.~P\'aramos, {\it Phys. Rev. D} {\bf 77}, 084018 (2008).
%
\bibitem{Wi} 
C.M.~Will, {\it Theory and Experiment in Gravitational Physics, Revised Ed.}
(Cambridge University Press, Cambridge, 1993).
%
\bibitem{Stix}
M.~Stix, {\it The Sun - An Introduction}
(Springer, Berlin, 2002).
%
\bibitem{Voyag}
D.A.Gurnett and W.S.~Kurth, {\it Nature Astron.} {\bf 3}, 1024 (2019).
%
\bibitem{DS}
A-C.~Davies, E.A.~Lim, J.~Sakstein and D.J.~Shaw,
{\it Phys. Rev. D} {\bf 85}, 123006 (2012).
%
\bibitem{Christ}
J.~Christensen-Dalsgaard, D.O.~Gough and M.J.~Thompson, {\it Astrophys. J.} {\bf 378}, 413 (1991).
%
\bibitem{Mariska} J.T.~Mariska, {\it Ann. Rev. Astron. Astrophys.} {\bf 24}, 23 (1986).
%
\bibitem{Jordan76} C.~Jordan, {\it Phil. Trans. R. Soc. Lond. A} {281}, 391 (1976).
%
\bibitem{Jordan} C.~Jordan {\it Astron. Astrophys.} {\bf 86}, 355 (1980).
%
\bibitem{RayDoyle} J.C.~Raymond and J.G.~Doyle,  {\it Astrophys. J.} {\bf 247}, 686 (1981).
%
\bibitem{Doschek} G.A.~Doschek, U.~Feldman, J.M.~Laming, H.P.~Warren, U.~Sch\"uhle and K.~Wilhelm,
{\it Astrophys. J.} {\bf 507}, 991 (1998).
%
\bibitem{Muller-Cyr} D.~M\"uller {\it et al}., {\it Astron. Astrophys.} {\bf 642}, A1 (2020).
%
\bibitem{Fox-Velli} N.J.~Fox {\it et al}., {\it Space Sci. Rev.} {\bf 204}, 7 (2016).
%
\bibitem{Tan-Han}
E.~Tandberg-Hanssen, {\it Solar Activity} (Blaisdell Publishing Company, Waltham, Massachusetts, 1967).
%
\bibitem{laneemdenyukawa} O.~Bertolami and J.~P\'aramos, {\it Phys. Rev. D} {\bf 71}, 023521 (2005).
%
\bibitem{Mullan}
D.J.~Mullan, {\it Physics of the Sun: A First Course} (CRC Press, Boca Raton, Florida, 2010).
%
\bibitem{Bahcall} J.N.~Bahcall, A.M.~Serenelli and S.~Basu, {\it Astrophys. J. Suppl.} {\bf 165}, 400 (2006).
%
\bibitem{DeJager} C.~De Jager, {\it Bull. Astron. Institut. Netherlands} {\bf 17}, 209 (1962).
%
\bibitem{Vernazza} J.E.~Vernazza, E.H.~Avrett and R. Loeser, {\it Astrophys. J. Suppl.} {\bf 45}, 635 (1981).
%
\bibitem{Kontar} E.P.~Kontar, I.G.~Hannah, N.L.S.~Jeffrey and M.~Battaglia, {\it Astrophys. J.} {\bf 717}, 250 (2010).
%
\bibitem{Mariska-book} J.T.~Mariska, {\it The Solar Transition Region} (Cambridge University Press, New York, 1992).
%
\bibitem{Parker} E.N.~Parker, {\it Astrophys. J.} {\bf 128}, 664 (1958).
%
\bibitem{Schwenn} R.~Schwenn, {\it JPL Solar Wind Five} (SEE N84-13067 03-92), 489 (1983).
%
\bibitem{Allen} R.C.~Allen {\it et al}., {\it Astrophys. J. Suppl.} {\bf 246}, 36 (2020).
%
\end{thebibliography}
\end{document}